\documentstyle[12pt,epsf]{article}
\begin{document}
\baselineskip 0.7cm

\begin{titlepage}

  \begin{flushright}
    TIT-HEP-304, NSF-ITP-95-127
    \\
    KEK-TH-450, LBL-37816
    \\
    UT-727, TU-491
    \\
    hep-ph/9510309
    \\
    October, 1995
  \end{flushright}

\begin{center}
  {\Large \bf Lepton-Flavor Violation\\
via Right-Handed Neutrino Yukawa Couplings \\
in Supersymmetric Standard Model  }
\vskip 0.5in
{\large
    J.~Hisano$^{(a,b)}$, T.~Moroi$^{(c,d)}$, K.~Tobe$^{(e,f)}$ and
    M.~Yamaguchi$^{(f)}$}
\vskip 0.4cm
{\it (a) Tokyo Institute of
    Technology, Department of Physics \\ Oh-okayama, Meguro, Tokyo
    152, Japan} \\
{\it (b) Institute for Theoretical Physics, University of California,\\
    Santa Barbara, CA 93106, U.S.A.}\\
{\it (c) Theory Group, KEK, Ibaraki 305, Japan } \\
{\it (d) Theoretical Physics Group, Lawrence Berkeley Laboratory, \\
University of California, Berkeley, CA 94720, U.S.A.}
\\
{\it (e) Department of Physics, University of Tokyo, Tokyo 113, Japan}
\\
{\it (f) Department of Physics, Tohoku University, Sendai 980-77, Japan}

\vskip 0.5in

\abstract
{Various lepton-flavor violating (LFV) processes in the supersymmetric
standard model with right-handed neutrino supermultiplets are
investigated in detail. It is shown that large LFV rates are obtained
when $\tan \beta $ is large. In the case where the mixing matrix in
the lepton sector has a similar structure as the Kobayashi-Maskawa
matrix and the third-generation Yukawa coupling is as large as that of
the top quark, the branching ratios can be as large as
$Br(\mu\rightarrow e\gamma)\simeq 10^{-11}$ and
$Br(\tau\rightarrow\mu\gamma)\simeq 10^{-7}$, which are within the
reach of future experiments. If we assume a large mixing angle
solution to the atmospheric neutrino problem, rate for the process
$\tau\rightarrow\mu\gamma$ becomes larger. We also discuss the
difference between our case and the case of the minimal $SU(5)$ grand
unified theory.}

\end{center}
\end{titlepage}

\section{Introduction}

Lepton-flavor violation (LFV), if observed in a future experiment, is
an evidence of new physics beyond the standard model, because the
lepton-flavor number is conserved in the standard model.
Since the processes do not suffer from a large ambiguity due to the
hadronic matrix elements, detailed analysis of the LFV processes will
reveal some properties of the high-energy physics.

One of the minimal extensions of the standard model with LFV is the
model with non-vanishing neutrino masses.  If the masses of the
neutrinos are induced by the seesaw mechanism~\cite{seesaw}, one has a
new set of Yukawa couplings involving the right-handed neutrinos.
Introduction of the new Yukawa couplings generally gives rise to the
flavor violation in the lepton sector, similar to its quark sector
counterparts.  In non-supersymmetric standard models, however, the
amplitudes of the LFV processes are proportional to inverse powers of
the right-handed neutrino mass scale which is typically much higher
than the electroweak scale, and as a consequence such rates are highly
suppressed.

If the model is supersymmetrized, the situation becomes quite
different.  LFV in the right-handed neutrino Yukawa couplings leads to
LFV in slepton masses through renormalization-group
effects~\cite{PRL57-961}. Then the LFV processes are only suppressed
by powers of supersymmetry (SUSY) breaking scale which is assumed to
be at the electroweak scale. Especially, in a previous
paper~\cite{tu476}, we pointed out that a large left-right mixing of
the slepton masses greatly enhances the rates for the LFV processes
such as $\mu\rightarrow e\gamma$ and $\tau\rightarrow\mu\gamma$. Due
to this effect, they can be within the reach of near future
experiments even if the mixing angle of the lepton sector is as small
as that of the quark sector.

In this paper, we will extend the previous analysis.  We are
interested in the following processes, \begin{itemize} \item $\mu
\rightarrow e \gamma$, \item $\tau \rightarrow \mu \gamma$, \item $\mu
\rightarrow e e e$, \item $\mu$-$ e $ conversion in nuclei,
\end{itemize} and calculate formulas for the interaction rates of the
above processes. In our calculation, we fully incorporate the mixing
of the slepton masses as well as the mixings in the neutralino and
chargino sectors.  Also the lepton Yukawa couplings in
higgsino-lepton-slepton vertices are retained, which yield another
type of enhanced diagrams in the large $\tan \beta $ region.  Then we
will discuss how large the interaction rates can be, assuming the
radiative electroweak symmetry breaking scenario~\cite{PTP68-927}.  We
find that a large value of $\tan\beta$ is realized with relatively
light superparticle mass spectrum, and thus the interaction rates can
indeed be enhanced. For the right-handed neutrino sector, we will
mainly consider the case where the Yukawa couplings of the
right-handed neutrinos are similar to those of the up-type quarks. We
will also discuss the case of large mixing between the second and
third generations, suggested by atmospheric neutrino problem.  In our
numerical analysis, we impose the constraints from the negative
searches for the SUSY particles, as well as the constraint from the
muon anomalous-magnetic dipole-moment $g-2$ to which superparticle
loops give non-negligible contributions especially in the large
$\tan\beta$ region.

The organization of our paper is as follows.  In the subsequent
section, we will review LFV in slepton masses in the presence of
the right-handed neutrinos.  In Section 3, we will give formulas of
the interaction rates of the various LFV processes. Results of our
numerical study are given in Section 4.  In Section 5, after
summarizing our results, we will compare our case with the case of the
$SU(5)$ grand unification briefly.  Renormalization-group equations
relevant to our analysis are summarized in Appendix A. In Appendix B,
we describe the interactions among neutralino
(chargino)-fermion-sfermion.  In Appendix C, we will give formulas of
the SUSY contribution to $g-2$.

\section{LFV in scalar lepton masses}

Throughout this paper, we consider the minimal SUSY standard model (MSSM)
plus three generation right-handed neutrinos. In this case, the
superpotential is given by
\newcommand{\mup}{m_{\tilde{u}ij}}
\newcommand{\md}{m_{\tilde{d}ij}}
\newcommand{\me}{m_{\tilde{e}ij}}
\newcommand{\mnu}{m_{\tilde{\nu}ij}}
\newcommand{\mq}{m_{\tilde{Q}ij}}
\newcommand{\ml}{m_{\tilde{L}ij}}
\newcommand{\mhd}{m_{\tilde{h1}ij}}
\newcommand{\mhu}{m_{\tilde{h2}ij}}
\newcommand{\sq}{{\tilde{q}_L}}
\newcommand{\su}{{\tilde{u}_R}}
\newcommand{\sd}{{\tilde{d}_R}}
\newcommand{\slep}{{\tilde{l}_L}}
\newcommand{\se}{{\tilde{e}_R}}
\newcommand{\snu}{{\tilde{\nu}_R}}
\begin{eqnarray}
\label{superpotential}
W&=&f_l^{ij}\epsilon_{\alpha \beta} H_1^{\alpha} E_i^c  L_j^{\beta}
+ f_{\nu}^{ij}\epsilon_{\alpha \beta} H_2^{\alpha} N_i^c L_j^{\beta}
+ f_{d}^{ij} \epsilon_{\alpha \beta} H_1^{\alpha} D_i^c  Q_j^{\beta}
+ f_{u}^{ij} \epsilon_{\alpha \beta} H_2^{\alpha} U_i^c Q_j^{\beta}
\nonumber \\
&&+ \mu \epsilon_{\alpha \beta} H_1^{\alpha} H_2^{\beta}
+ \frac{1}{2} M_{\nu}^{ij} N_i^c N_j^c,
\end{eqnarray}
where $L_i$ represents the chiral multiplet of a $SU(2)_L$ doublet
lepton, $E_i^c$ a $SU(2)_L$ singlet charged lepton, $N_i^c$ a
right-handed neutrino which is singlet under the standard-model gauge
group, $H_1$ and $H_2$ two Higgs doublets with opposite hypercharge.
Similarly $Q$, $U$ and $D$ represent chiral multiplets of quarks of a
$SU(2)_L$ doublet and two singlets with different $U(1)_Y$ charges.
Three generations of leptons and quarks are assumed and thus the
subscripts $i$ and $j$ run over 1 to 3. The symbol $\epsilon_{\alpha
\beta}$ is an anti-symmetric tensor with $\epsilon_{12}=1$.  The
Yukawa interactions are derived from the superpotential via
\begin{equation}
{\cal{L}}=+\frac{1}{2}\sum_{i,j} \frac{\partial^2 W}
{\partial \phi_i \partial \phi_j} \psi_i \psi_j + h.c..
\nonumber
\end{equation}

SUSY is softly broken in our model.  The general soft SUSY breaking
terms are given as
\begin{eqnarray}
\label{softbreaking}
-{\cal{L}}_{soft}&=&(m_{\tilde Q}^2)_i^j {\tilde q}_{L}^{\dagger i}
{\tilde q}_{Lj}
+(m_{\tilde u}^2)^i_j {\tilde u}_{Ri}^* {\tilde u}_{R}^j
+(m_{\tilde d}^2)^i_j {\tilde d}_{Ri}^* {\tilde d}_{R}^j
\nonumber \\
& &+(m_{\tilde L}^2)_i^j {\tilde l}_{L}^{\dagger i}{\tilde l}_{Lj}
+(m_{\tilde e}^2)^i_j {\tilde e}_{Ri}^* {\tilde e}_{R}^j
+(m_{\tilde \nu}^2)^i_j {\tilde \nu}_{Ri}^* {\tilde \nu}_{R}^j
\nonumber \\
& &+{\tilde m}^2_{h1}h_1^{\dagger} h_1
+{\tilde m}^2_{h2}h_2^{\dagger} h_2
+(B \mu h_1 h_2
+ \frac{1}{2}B_{\nu}^{ij} M_{\nu}^{ij}
 {\tilde \nu}_{Ri}^* {\tilde \nu}_{Rj}^* + h.c.)
\nonumber \\
& &+ ( A_d^{ij}h_1 {\tilde d}_{Ri}^*{\tilde q}_{Lj}
+A_u^{ij}h_2 {\tilde u}_{Ri}^*{\tilde q}_{Lj}
+A_l^{ij}h_1 {\tilde e}_{Ri}^*{\tilde l}_{Lj}
+A_{\nu}^{ij}h_2 {\tilde \nu}_{Ri}^* {\tilde l}_{Lj}
\nonumber \\
& & +\frac{1}{2}M_1 {\tilde B}_L^0 {\tilde B}_L^0
+\frac{1}{2}M_2 {\tilde W}_L^a {\tilde W}_L^a
+\frac{1}{2}M_3 {\tilde G}^a {\tilde G}^a +h.c.).
\end{eqnarray}
Here the first four lines are  soft terms for  sleptons, squarks
and the Higgs bosons,  while the last line gives gaugino mass terms.

We now discuss LFV in the Yukawa couplings.  Suppose that the Yukawa
coupling matrix $f_l^{ij}$ and the mass matrix of the right-handed
neutrinos $M_{\nu}^{ij}$ are diagonalized as $ f_{l i} \delta^{ij}$
and $M_{Ri} \delta^{ij}$, respectively.\footnote
{We can always choose $f_l^{ij}$ and $M_{\nu}^{ij}$ to be diagonal by
using unitary transformations of $L$, $E^c$ and $N^c$.}
Then, in this basis, the neutrino Yukawa couplings $f_{\nu}^{ij}$ are
not generally diagonal, giving rise to LFV.  An immediate consequence
is neutrino oscillation.  Writing $f_{\nu}^{ij}=U^{ik} f_{\nu k}
V^{kj}$ with $U$, $V$ unitary matrices, we obtain the neutrino mass
matrix induced by the seesaw mechanism
\begin{eqnarray}
m_{\nu}
& =& f_{\nu}^{{\sf T}} M^{-1}_{\nu} f_{\nu}  \times
  \frac{v^2}{2} \sin^2 \beta
\nonumber \\
& =& V^{{\sf T}}
    \left( \begin{array}{ccc}
     f_{\nu 1} &   &   \\
       & f_{\nu 2} &   \\
       & & f_{\nu 3}
   \end{array} \right)
   U^{{\sf T}}
   \left( \begin{array}{ccc}
     \frac{1}{M_{R 1}} & & \\
     & \frac{1}{M_{R 2}} & \\
     & & \frac{1}{M_{R 3}}
   \end{array} \right)
   U
   \left( \begin{array}{ccc}
     f_{\nu 1} &   &   \\
       & f_{\nu 2 } &   \\
       & & f_{\nu 3}
   \end{array} \right) V
\nonumber \\
& &  \times  \frac{v^2}{2} \sin^2 \beta,
\end{eqnarray}
where $\frac{1}{2}v^2=\langle h_1\rangle^2+\langle h_2\rangle^2\simeq
(174{\rm GeV})^2$ and $\tan\beta=\langle h_2\rangle /\langle
h_1\rangle$. (Here, $\langle\cdots\rangle$ stands for the
vacuum expectation value of the quantity.)  Throughout this paper, we
assume that $M_{\nu}$ is proportional to the unit matrix $M_{\nu}^{
ij}=M_R\delta^{ij}$, for simplicity.  Then, if we disregard possible
complex phases in $U$, the above can be rewritten as
\begin{equation}
  m_{\nu} = \frac{1}{M_R} V^{{\sf T} }
   \left(
   \begin{array}{ccc}
      f_{\nu 1}^2 & & \\
       &  f_{\nu 2}^2 & \\
       & & f_{\nu 3}^2
   \end{array} \right)
   V \times \frac{v^2}{2} \sin ^2 \beta.
\end{equation}
Thus as far as $V \neq 1$ and the mass eigenvalues are non-degenerate,
we have neutrino oscillation which is a target of current and future
experiments.

The smallness of the neutrino masses implies that the scale $M_R$ is
very high, $\sim 10^{12}$ GeV or even higher.  In the standard model
with right-handed neutrinos, the flavor violating processes such as
$\mu \rightarrow e\gamma$, $\tau\rightarrow\mu\gamma$ {\it etc.},
whose rates are proportional to inverse powers of $M_R$, would be
highly suppressed with such a large $M_R$ scale, and hence those would
never be seen experimentally.

However, if there exists SUSY broken at the electroweak scale, we may
expect that the rates of these LFV processes will be much larger than
the non-supersymmetric case.  The point is that the lepton-flavor
conservation is not a consequence of the standard-model gauge symmetry
and renormalizability in the supersymmetric case, even in the absence
of the right-handed neutrinos. Indeed, slepton mass terms can violate
the lepton-flavor conservation in a manner consistent with the gauge
symmetry.  Thus the scale of LFV can be identified with the
electroweak scale, much lower than the right-handed neutrino scale
$M_R$.  However, an order-of-unity violation of the lepton-flavor
conservation at the electroweak scale would cause disastrously large
rates for $\mu\rightarrow e\gamma$ and others.  Also, arbitrary squark
masses result in too large rates for various
flavor-changing-neutral-current processes involving squark loops.  To
avoid these problems, one often considers that the sleptons and the
squarks are degenerate in masses among those with the same gauge
quantum numbers in the tree-level Lagrangian at a certain
renormalization scale. In the following, we will assume a somewhat
stronger hypothesis that all SUSY breaking scalar masses are universal
at the gravitational scale $M \equiv m_{pl}/\sqrt{8 \pi} \sim 2\times
10^{18}$GeV, {\it i.e.}, we adopt the minimal supergravity type
boundary conditions. Thus we will consider the following type of soft
terms,
\begin{itemize}
\item universal scalar mass ($m_0$),\\
  all scalar masses of the type $(m_{\tilde f}^2)_i^j$ and $\tilde
  m^2_{h_i}$ ($i=1,2$) take common value $m_0^2$,
\item universal $A$-parameter,\\
  $A_{f}^{ij}=a f_{f}^{ij} m_0$ with $a$ being a  constant of order unity,
\end{itemize}
at the renormalization scale $M$.\footnote
{In fact, there is another SUSY breaking parameter $B$, which gives a
mixing term of the two Higgs bosons $h_1$ and $h_2$. For a given value
of $\tan\beta$, we fix this parameter $B$ (and also the SUSY invariant
Higgs mass $\mu$) so that the Higgs bosons have correct vacuum
expectation values, $\langle h_1 \rangle =v\cos\beta/\sqrt2$ and
$\langle h_2 \rangle=v\sin\beta/\sqrt2$.}
As for the gaugino masses, for simplicity, we choose the boundary
condition so that they satisfy the so-called grand unified theory
(GUT) relation at low energies. Note that the universal scalar masses
are given in a certain class of supergravity models with hidden sector
SUSY breaking~\cite{Nilles}.  Those soft SUSY breaking terms suffer
from renormalization via gauge and Yukawa interactions, which can be
conveniently expressed in terms of the renormalization-group equations
(RGEs).  The RGEs relevant in our analysis will be given in Appendix
A.  An important point is that, through this renormalization effect,
LFV in the Yukawa couplings induces LFV in the slepton masses at low
energies even if the scalar masses are universal at high energy. Due
to this fact, lepton-flavor conservation is violated at low energies.

We can solve the RGEs numerically with the boundary conditions given
above.  It is, however, instructive to consider here a simple
approximation to estimate the LFV contribution to the slepton masses.
Since the $SU(2)_L$ doublet lepton multiplets have the lepton-flavor
violating Yukawa couplings with the right-handed neutrino multiplets,
the LFV effect most directly appears in the mass matrix of the doublet
sleptons.  The RGEs for them can be written as (see Appendix~A)
\begin{eqnarray}
  \mu\frac{d}{d \mu}(m^2_{\tilde L})_i^j
 &=& \left( \mu\frac{d}{d \mu}(m^2_{\tilde L})_i^j \right)_{\rm MSSM}
\nonumber \\
& & +\frac{1}{16 \pi^2}
  \left[ (m^2_{\tilde L} f_{\nu}^{\dagger}f_{\nu}+
  f_{\nu}^{\dagger}f_{\nu} m^2_{\tilde L})_i^j
  +2 (f^{\dagger}_{\nu} m_{\tilde \nu}^2 f_{\nu}
     + \tilde m_{h2}^2 f^{\dagger}_{\nu} f_{\nu}
     +A_{\nu}^{\dagger} A_{\nu})_i^j \right].
\label{RGE-lslepton}
\end{eqnarray}
Here $(\mu \frac{d}{d \mu}(m^2_{\tilde L})_i^j)_{\rm MSSM}$ denotes the
RGE in case of the MSSM, and the terms explicitly written are
additional contributions by  the
right-handed neutrino Yukawa couplings. An iteration gives an approximate
solution for the additional contributions to the mass terms
\begin{eqnarray}
  (\Delta m_{\tilde L}^2)_i^j
& \approx &
  -\frac{\ln (M/M_R)}{16 \pi^2}
   \left( 6 m_0^2 (f_{\nu}^{\dagger} f_{\nu})_i^j
         +2 (A_{\nu}^{\dagger} A_{\nu})_i^j \right)
\nonumber \\
& = &
    -\frac{\ln (M/M_R)}{16 \pi^2}
   ( 6+2a^2) m_0^2 (f_{\nu}^{\dagger} f_{\nu})_i^j ,
\label{slepton-approx}
\end{eqnarray}
where we have used the universal scalar mass and A-parameter conditions.
  In Eq.~(\ref{slepton-approx}),
\begin{equation}
   (f_{\nu}^{\dagger} f_{\nu})_i^j
 = f^{\dagger}_{\nu ik} f_{\nu}^{kj}
 = V^{*}_{ki} |f_{\nu k}|^2 V^{kj},
\end{equation}
so that the slepton mass $(m_{\tilde L}^2)_i^j$ indeed has the
generation mixing if $V$ differs from the unit matrix in the basis
that the charged lepton Yukawa coupling $f_l$ are diagonal.

Lack of our knowledge on the neutrino Yukawa couplings prevents us
from giving a definite prediction of the slepton mass matrix, and thus
the rates of the LFV processes.  Nevertheless, it is important to study
how large the interaction rates for the LFV processes can be for some
typical cases and to see whether those signals can be tested by
experiments.  In this paper, we shall consider the following typical
two cases: {\em case 1\/}) the mixing matrix $V$ is identical to the
Kobayashi-Maskawa (KM) matrix in the quark sector $V_{KM}$, and {\em
case 2\/}) the mixing matrix is given so that it can explain
atmospheric neutrino deficit by the large-mixing
$\nu_{\tau}$-$\nu_{\mu}$ oscillation.  In the latter case, we only
consider $\tau\rightarrow\mu\gamma$, the generation mixing between
the second and third ones.

\section{Interaction rates for LFV processes}

In this section we give formulas of the interaction rates for the
LFV processes we consider.  Results of our
numerical calculation will be given in the next section.

We first explain how the rates for $\mu \rightarrow e \gamma$ and
$\tau \rightarrow \mu \gamma$ can be enhanced compared with the naive
expectation when $\tan \beta$ is large.  Here, we consider in the basis
where the neutralino/chargino interactions to the leptons and the
sleptons are flavor diagonal and the effect of the flavor violation in
the lepton sector is involved by the mass insertions $(m^{2}_{\tilde
L})^j_i$ $(i \neq j)$.  First, let us consider contribution from winos
and bino, the $SU(2)_L \times U(1)_Y$ gauginos, neglecting the mixing
in the chargino/neutralino sector.  A naive estimate on the branching
ratio yields
\begin{equation}
Br(l_j \rightarrow l_i \gamma ) \propto \frac{\alpha^3}{G_F^2}
       \frac{((m_{\tilde L}^2)_i^j)^2}{m_S^8},
\end{equation}
where $m_S$ is the typical mass of superparticles, $\alpha$ the fine
structure constant and $G_F$ the Fermi constant. The contribution from
Feynman diagrams Fig.~1(a) and (b) follows this estimate.  However, as
emphasized in our previous paper~\cite{tu476}, the diagram Fig.~1(c)
which picks up the left-right mixing of the sleptons and exchanges
the bino in the loop can give much larger contribution, when $\mu \tan
\beta$ is much larger than the masses of the other superparticles. Indeed we
estimate the ratio of the amplitudes
\begin{equation}
 \frac{ Amp. (1.c)}{Amp. (1.a+b)}
 \sim
 \frac{M_1 m^2_{LR jj}}{m_{l_j} m_S^2}
 \sim
 \frac{\mu \tan \beta}{m_S} \frac{M_1}{m_S},
\end{equation}
with $m_{l_j}$ being the charged lepton $l_j$ mass.  In
Ref.~\cite{tu476}, we numerically showed that this enhancement really
occurs for the case of large $\mu\tan\beta$.

If we take account of the gaugino-higgsino mixing in the
chargino/neutralino sector, we find another type of diagram which
enhances the amplitude when $\tan \beta$ is large but $\mu $ is
comparable to the masses of the other superparticles.  It is shown in
Fig.~2.  In this diagram, one has the mixing between the higgsino and
the gaugino which is proportional to $v \sin \beta$, the vacuum
expectation value of $h_2$, and involves the Yukawa coupling of
higgsino-lepton-slepton, $f_{l_j}=-\sqrt{2} m_{l_j}/(v \cos \beta)$.
The sleptons inside a loop are left-handed ones.  Thus the amplitude
is proportional to $\tan\beta$, and
\begin{equation}
  \frac{Amp. (2)}{Amp. (1.a+b)} \sim \tan \beta.
\end{equation}
  Note that this type of diagram includes
neutralino-exchange graphs as well as a chargino-exchange graph.


In this work, we are interested in the following LFV processes; $\mu
\rightarrow e \gamma$ and $\tau \rightarrow \mu \gamma$, $\mu^-
\rightarrow e^- e^- e^+$ and $\mu$-$e$ conversion in nuclei.
To obtain the interaction rates for these processes, we perform full
diagonalization of the slepton mass matrices numerically and consider
mixing in the chargino and neutralino sectors.

We write the interaction Lagrangian of fermion-sfermion-neutralino as
\begin{equation}
{\cal{L}}=\bar{f}_i (N_{iAX}^{R(f)} P_R + N_{iAX}^{L(f)} P_L) \tilde{\chi}^0_A
\tilde{f}_X + h.c..
\end{equation}
In this section, $f_i$ ($f=l,\nu,d,u$) represents a fermion in {\em
mass eigenstate} with the generation index $i$ ($i=1,2,3$), and
$\tilde f_X$ a sfermion in mass eigenstate. The subscript $X$ runs
from 1 to 3 for $\tilde \nu$ and from 1 to 6 for the other sfermions,
$\tilde l,\tilde d$ and $\tilde u$.  A neutralino mass eigenstate is
denoted by $\tilde \chi^0_A$ ($A=1, \cdots, 4$) and
$P_{R,L}=\frac{1}{2}(1\pm\gamma_5)$.  The coefficients
$N_{iAX}^{R(f)}$ and $N_{iAX}^{L(f)}$ depend on the mixing matrices of
the neutralino sector and of the sfermions.  Their explicit forms will
be given in Appendix B.  Similarly the fermion-sfermion-chargino
interaction is written as
\begin{eqnarray}
{\cal L}&=&\bar{l}_i (C_{iAX}^{R(l)} P_R + C_{iAX}^{L(l)} P_L)
\tilde{\chi}^-_A
\tilde{\nu}_X
\nonumber \\
&&+\bar{\nu}_i (C_{iAX}^{R(\nu) } P_R + C_{iAX}^{L (\nu) } P_L)
\tilde{\chi}^+_A
\tilde{l}_X
\nonumber \\
&&+\bar{d}_i (C_{iAX}^{R(d)} P_R + C_{iAX}^{L(d)} P_L)
\tilde{\chi}^-_A
\tilde{u}_X
\nonumber \\
&&+\bar{u}_i (C_{iAX}^{R(u)} P_R + C_{iAX}^{L(u)} P_L)
\tilde{\chi}^+_A
\tilde{d}_X + h.c.,
\end{eqnarray}
where $\tilde \chi^-_A$ ($A=1,2$) is a chargino mass eigenstate.  The
explicit forms of the coefficients can also be found in the Appendix~B.

\subsection{Effective Lagrangian for LFV processes}
As a first step to compute the LFV rates, let us write down the
effective interactions (or amplitudes) relevant for our purpose.

\subsubsection{$l^-_j \rightarrow l^-_i \gamma^*$}
The off-shell amplitude for $l^-_j\rightarrow l^-_i \gamma^*$ is
generally written as
\begin{eqnarray}
T=e \epsilon^{\alpha*} \bar{u}_i (p-q)
\left [
q^2 \gamma_\alpha (A_1^L P_L + A_1^R P_R )
+m_{l_j} i \sigma_{\alpha \beta} q^\beta (A_2^L P_L + A_2^R P_R)
\right]
u_j(p),
\label{Penguin}
\end{eqnarray}
in the limit of $q\rightarrow 0$ with $q$ being the photon momentum.
Here, $e$ is the electric charge, $\epsilon^*$ the photon polarization
vector, $u_i$ (and $v_i$ in the expressions below) the wave function
for (anti-) lepton, and $p$ the momentum of the particle $l_j$.  In
the present case, the Feynman diagrams contributing to the above
amplitude are depicted by Fig.~3.  Each coefficients in the above can
be written as a sum of the two terms,
\begin{eqnarray}
A_a^{L, R}&=&A_a^{(n)L, R}+A_a^{(c)L, R}~~~~(a=1,2),
\nonumber
\end{eqnarray}
where $ A_a^{(n)L, R}$ and $A_a^{(c)L, R}$ stand for the contributions
from the neutralino loops and from the chargino loops, respectively.
We calculate them and find that the neutralino contributions are given
by
\begin{eqnarray}
A_1^{(n)L}&=&\frac{1}{576 \pi^2} N_{iAX}^{R(l)} N_{jAX}^{R(l)*}
\frac{1}{m^2_{\tilde{l}_X}} \frac{1}{(1-x_{AX})^4}
\nonumber \\
&&\times
(2-9x_{AX}+18x_{AX}^2-11x_{AX}^3 +6x_{AX}^3 \ln x_{AX}),
\\
A_2^{(n)L}&=&\frac{1}{32 \pi^2}\frac{1}{m^2_{\tilde{l}_X}}
\left[
 N_{iAX}^{L(l)} N_{jAX}^{L(l)*}
\frac{1}{6 (1-x_{AX})^4}
\right.
\nonumber \\
&&\times
(1-6x_{AX}+3x_{AX}^2+2x_{AX}^3-6x_{AX}^2\ln x_{AX})
\nonumber \\
&&\left.
+N_{iAX}^{L(l)} N_{jAX}^{R(l)*} \frac{M_{\tilde{\chi}_A^0}}{m_{l_j}}
\frac{1}{(1-x_{AX})^3}
(1-x_{AX}^2+2x_{AX} \ln x_{AX})
\right],
\\
A_a^{(n)R}&=&A_a^{(n)L}|_{L \leftrightarrow R}~~~~(a=1,2),
\end{eqnarray}
where $x_{AX}=M^2_{\tilde{\chi}^0_A}/m^2_{\tilde{l}_X}$ is the ratio
of the neutralino mass squared, $M^2_{\tilde{\chi}^0_A}$, to the
charged slepton mass squared, $m^2_{\tilde l_X}$. (Summation over the
indices $A$ and $X$ are assumed to be understood.)  The chargino
contributions are
\begin{eqnarray}
A_1^{(c)L}&=&-\frac{1}{576 \pi^2}
C_{iAX}^{R(l)} C_{jAX}^{R(l)*}\frac{1}{m^2_{\tilde{\nu}_X}}
\frac{1}{(1-x_{AX})^4}
\nonumber \\
&&\times \left \{
16-45x_{AX}+36x_{AX}^2-7x_{AX}^3+6(2-3x_{AX}) \ln x_{AX}
\right \},
\\
A_2^{(c)L}&=&-\frac{1}{32 \pi^2}\frac{1}{m^2_{\tilde{\nu}_X}}
\left[
 C_{iAX}^{L(l)} C_{jAX}^{L(l)*}
\frac{1}{6 (1-x_{AX})^4}
\right.
\nonumber \\
&&\times
(2+3x_{AX}-6x_{AX}^2+x_{AX}^3+6x_{AX} \ln x_{AX})
\nonumber \\
&&\left.
+C_{iAX}^{L(l)} C_{jAX}^{R(l)*} \frac{M_{\tilde{\chi}_A^-}}{m_{l_j}}
\frac{1}{(1-x_{AX})^3}
(-3+4x_{AX}-x_{AX}^2-2 \ln x_{AX})
\right],
\\
A_a^{(c)R}&=&A_a^{(c)L}|_{L \leftrightarrow R}~~~~(a=1,2).
\end{eqnarray}
Here, $x_{AX}=M^2_{\tilde{\chi}^-_A}/m^2_{\tilde{\nu}_X}$, where
$M_{\tilde{\chi}^-_A}$ and $m_{\tilde \nu_X}$ are the masses for the
chargino $\tilde \chi^-_A$ and the sneutrino ${\tilde \nu_X}$,
respectively.

\subsubsection{
$l^-_j \rightarrow l^-_i~l^-_i~l^+_i$}

We next consider the process $l^-_j \rightarrow l^-_i~l^-_i~l^+_i$
(including $\mu^- \rightarrow e^- e^- e^+$).  The effective amplitude
consists of the contributions from the Penguin-type diagrams and from
the box-type diagrams.  The former contribution can be computed using
Eq.~(\ref{Penguin}),  with the result
\begin{eqnarray}
T_{\gamma-penguin}&=& \bar{u}_{i}(p_1)
\left[
q^2 \gamma_{\alpha}(A_1^L P_L + A_1^R P_R) +
m_{l_j} i \sigma_{\alpha \beta}q^\beta(A_2^L P_L + A_2^R P_R)
\right]
u_{j}(p)
\nonumber \\
&& \times \frac{e^2}{q^2}\bar{u}_{i}(p_2) \gamma^{\alpha} v_{i}(p_3)
 - (p_1 {\leftrightarrow} p_2).
\label{Penguin3e}
\end{eqnarray}
Furthermore, there are the other Penguin-type diagrams in which the
Z boson is exchanged as shown in Fig.~4. This amplitude is
\begin{eqnarray}
T_{Z-penguin}= \frac{g_Z^2}{m_{Z}^2} \bar{u}_{i}(p_1) \gamma^{\mu}
(F_L P_L + F_R P_R) u_{j}(p) \bar{u}_i (p_2) \gamma^{\mu}
(Z_L^l P_L + Z_R^l P_R) v_i(p_3)
\nonumber \\
 - (p_1 {\leftrightarrow} p_2),
\label{Z_penguin}
\end{eqnarray}
where $F_{L(R)}=F_{L(R)}^{(c)}+F_{L(R)}^{(n)}$. The chargino
contribution $F_{L(R)}^{(c)}$ and the neutralino contribution
$F_{L(R)}^{(n)}$ are\footnote
{The Penguin-type diagrams of Z boson contributing to the LFV events
do not necessarily need to have chirality flip of lepton as
$\mu\rightarrow e\gamma$.  Therefore, the diagrams picking up Yukawa
coupling of higgsino-fermion-sfermion can not become the dominant
contribution in Z boson Penguin-type diagrams and we neglect them in
the above equations.}
\begin{eqnarray}
F_L^{(c)}&=&-\frac{C_{iAX}^{R(l)}C_{jBX}^{R(l)*}}{16 \pi^2}
\left[
\frac{(O_R)_{A2}(O_R)_{B2}}{4} F_{(X,A,B)}
-\frac{(O_L)_{A2}(O_L)_{B2}}{2} G_{(X,A,B)}
\right],
\\
F_R^{(c)}&=& 0,
\\
F_L^{(n)}&=&\frac{N_{iAX}^{R(l)}N_{jBX}^{R(l)*}}{16 \pi^2}
\frac{(O_N)_{A3} (O_N)_{B3}-(O_N)_{A4} (O_N)_{B4}}{2}
\left( F_{(X,A,B)}+ 2 G_{(X,A,B)}\right),
\\
F_R^{(n)}&=&-F_L^{(n)}|_{L \leftrightarrow R}.
\end{eqnarray}
Here, $O_{L,R}$ and $O_N$ are orthogonal matrices to diagonalize the
mass matrices of the chargino and neutralino (see Appendix~B), and
$F_{(X,A,B)}$ and $G_{(X,A,B)}$ are given by
\begin{eqnarray}
F_{(X,A,B)}&=&
\ln x_{AX}  + \frac{1}{x_{AX}-x_{BX}}
\left(\frac{x_{AX}^2 \ln x_{AX}}{1-x_{AX}}
-\frac{x_{BX}^2 \ln x_{BX}}{1-x_{BX}}
\right),
\\
G_{(X,A,B)}&=&
\frac{M_{\chi_A}M_{\chi_B}}{m_{\tilde{l}_X}^2}
\frac{1}{x_{AX}-x_{BX}}
\left(\frac{x_{AX} \ln x_{AX}}{1-x_{AX}}
-\frac{x_{BX} \ln x_{BX}}{1-x_{BX}}
\right).
\end{eqnarray}
In these functions, $M_{\tilde{\chi}_A}$ and $m_{\tilde{l}_X}$
denote neutralino mass and charged slepton mass
in the neutralino contribution, and chargino mass and
sneutrino mass in the chargino contribution.
And, in Eq.~(\ref{Z_penguin}) the coefficient $Z_{L(R)}^l$ denotes
Z boson coupling to charged lepton $l_{L(R)}$, that is,
\begin{eqnarray}
Z_{L(R)}^l=T_{3L(R)}^l-Q^l_{em}\sin^2 \theta_W,
\label{Z_coupling}
\end{eqnarray}
where $T_{3L(R)}^l$ and $Q^l_{em}$ represent weak isospin
($T_{3L}^l=-\frac{1}{2},~~T_{3R}^l=0$) and electric charge
($Q^l_{em}=-1$) of $l_{L(R)}$ respectively.

The box-type Feynman diagrams are given in Fig.~5, and we can write
their amplitude as
\begin{eqnarray}
T_{box}&=& B_1^L e^2~ \bar{u}_i(p_1) \gamma^\alpha P_L u_{j}(p)
{}~\bar{u}_i(p_2) \gamma_{\alpha} P_L v_i(p_3)
 \nonumber \\
&&+ B_1^R e^2~ \bar{u}_i(p_1) \gamma^\alpha P_R u_{j}(p)
{}~\bar{u}_i(p_2) \gamma_{\alpha} P_R v_i(p_3)
\nonumber \\
&&+ B_2^L e^2
\left \{
\bar{u}_i(p_1) \gamma^\alpha P_L u_{j}(p)~
\bar{u}_i(p_2) \gamma_{\alpha} P_R v_i(p_3)
 - (p_1 {\leftrightarrow} p_2)
\right \}
\nonumber \\
&&+ B_2^R e^2
\left \{
\bar{u}_i(p_1) \gamma^\alpha P_R u_{j}(p)~
\bar{u}_i(p_2) \gamma_{\alpha} P_L v_i(p_3)
 - (p_1 {\leftrightarrow} p_2)
\right \}
\nonumber \\
&&+ B_3^L e^2
\left \{
\bar{u}_i(p_1) P_L u_{j}(p)~
\bar{u}_i(p_2) P_L v_i(p_3)
 - (p_1 {\leftrightarrow} p_2)
\right \}
\nonumber \\
&&+ B_3^R e^2
\left \{
\bar{u}_i(p_1) P_R u_{j}(p)~
\bar{u}_i(p_2) P_R v_i(p_3)
 - (p_1 {\leftrightarrow} p_2)
\right \}
\nonumber \\
&&+ B_4^L e^2
\left \{
\bar{u}_i(p_1) \sigma_{\mu \nu} P_L u_{j}(p)~
\bar{u}_i(p_2) \sigma^{\mu \nu} P_L v_{i}(p_3)
 - (p_1 {\leftrightarrow} p_2)
\right \}
\nonumber \\
&&+ B_4^R e^2
\left \{
\bar{u}_i(p_1) \sigma_{\mu \nu} P_R u_{j}(p)~
\bar{u}_i(p_2) \sigma^{\mu \nu} P_R v_{i}(p_3)
 - (p_1 {\leftrightarrow} p_2)
\right \},
\label{box3e}
\end{eqnarray}
where
\begin{eqnarray}
B^{L, R}_a=B^{(n)L, R}_a+B^{(c)L, R}_a~~~~(a=1, \cdots, 4).
\end{eqnarray}
The first term represents the neutralino contribution, which we find
to be
\begin{eqnarray}
e^2B_1^{(n)L}&=&\frac{1}{2} J_{4(A,B,X,Y)} N^{R(l)*}_{jAX}N^{R(l)}_{iAY}
N^{R(l)*}_{iBY} N^{R(l)}_{iBX}
\nonumber \\&&
+ I_{4(A,B,X,Y)} M_{\tilde{\chi}_A^0} M_{\tilde{\chi}_B^0}
 N^{R(l)*}_{jAX}N^{R(l)*}_{iAY}
N^{R(l)}_{iBY} N^{R(l)}_{iBX},
\\
e^2B_2^{(n)L}&=&\frac{1}{4}J_{4(A,B,X,Y)}
\left \{
 N^{R(l)*}_{jAX}N^{R(l)}_{iAY}
N^{L(l)*}_{iBY} N^{L(l)}_{iBX}+
 N^{R(l)*}_{jAX}N^{L(l)*}_{iAY}
N^{R(l)}_{iBY} N^{L(l)}_{iBX}
\right.
\nonumber \\
&&\left.
- N^{R(l)*}_{jAX}N^{L(l)*}_{iAY}
N^{L(l)}_{iBY} N^{R(l)}_{iBX}
\right \}
\nonumber \\&&
-\frac{1}{2} I_{4(A,B,X,Y)} M_{\tilde{\chi}_A^0} M_{\tilde{\chi}_B^0}
N^{R(l)*}_{jAX}N^{L(l)}_{iAY}
N^{L(l)*}_{iBY} N^{R(l)}_{iBX},
\\
e^2B_3^{(n)L}&=& I_{4(A,B,X,Y)} M_{\tilde{\chi}_A^0} M_{\tilde{\chi}_B^0}
\Big\{
N^{R(l)*}_{jAX} N^{L(l)}_{iAY}
N^{R(l)*}_{iBY} N^{L(l)}_{iBX}
\nonumber \\ &&
+\frac{1}{2} N^{R(l)*}_{jAX} N^{R(l)*}_{iAY}
N^{L(l)}_{iBY} N^{L(l)}_{iBX}
\Big\} ,
\\
e^2B_4^{(n)L}&=& \frac{1}{8}I_{4(A,B,X,Y)} M_{\tilde{\chi}_A^0}
M_{\tilde{\chi}_B^0}
N^{R(l)*}_{jAX} N^{R(l)*}_{iAY} N^{L(l)}_{iBY} N^{L(l)}_{iBX},
\\
B_a^{(n)R}&=&B_a^{(n)L}|_{L \leftrightarrow R} ~~~~~(a=1, \cdots, 4).
\end{eqnarray}
The chargino contribution is
\begin{eqnarray}
e^2B_1^{(c)L}&=&\frac{1}{2} J_{4(A,B,X,Y)} C^{R(l)*}_{jAX}C^{R(l)}_{iAY}
C^{R(l)*}_{iBY} C^{R(l)}_{iBX},
\\
e^2B_2^{(c)L}&=&\frac{1}{4}J_{4(A,B,X,Y)}
 C^{R(l)*}_{jAX}C^{R(l)}_{iAY}
C^{L(l)*}_{iBY} C^{L(l)}_{iBX}
\nonumber \\
&&
-\frac{1}{2} I_{4(A,B,X,Y)} M_{\tilde{\chi}_A^-} M_{\tilde{\chi}_B^-}
C^{R(l)*}_{jAX}C^{L(l)}_{iAY}
C^{L(l)*}_{iBY} C^{R(l)}_{iBX},
\\
e^2B_3^{(c)L}&=&I_{4(A,B,X,Y)} M_{\tilde{\chi}_A^-} M_{\tilde{\chi}_B^-}
C^{R(l)*}_{jAX} C^{L(l)}_{iAY}
C^{R(l)*}_{iBY} C^{L(l)}_{iBX},
\\
B_4^{(c)L}&=&0,
\\
B_a^{(c)R}&=&B_a^{(c)L}|_{L \leftrightarrow R}~~~~~(a=1, \cdots, 4),
\end{eqnarray}
where
\begin{eqnarray}
iJ_{4(A,B,X,Y)}&=&\int \frac{d^4k}{(2 \pi)^4}\frac{k^2}
{(k^2-M_{\tilde{\chi}_A}^{2})(k^2-M_{\tilde{\chi}_B}^{2})
(k^2-m_{\tilde{l}_X}^2)(k^2-m_{\tilde{l}_Y}^2)},
\\
iI_{4(A,B,X,Y)}&=&\int \frac{d^4k}{(2 \pi)^4}\frac{1}
{(k^2-M_{\tilde{\chi}_A}^{2})(k^2-M_{\tilde{\chi}_B}^{2})
(k^2-m_{\tilde{l}_X}^2)(k^2-m_{\tilde{l}_Y}^2)}.
\end{eqnarray}
Here, $M_{\tilde{\chi}_A}$ and $m_{\tilde{l}_X}$
denote neutralino mass and charged slepton mass
in the neutralino contribution, and chargino mass and
sneutrino mass in the chargino contribution.

\subsubsection{$\mu$-$e$ conversion in nuclei}
Finally, we give the formulas for the $\mu$-$e$ conversion in nuclei,
{\it i.e.}, the process ($\mu +(A,Z) \rightarrow e + (A,Z)$) where $Z$
and $A$ denote the proton and atomic numbers in a nucleus,
respectively.  The contribution again consists of the Penguin-type
diagrams and the box-type diagrams.  The box-type Feynman diagrams are
depicted in Fig.~6(b) and (c).  We give the effective Lagrangian
relevant to this process at the quark level.  We find that the
Penguin-type diagrams give the following terms,
\begin{eqnarray}
{\cal L}^{penguin}_{eff}&=&-\frac{e^2}{q^2}
\bar{e}
\left[
q^2 \gamma_{\alpha}(A_1^L P_L+A_1^R P_R)
 +m_{\mu}i\sigma_{\alpha \beta} q^{\beta}(A_2^L P_L+A_2^R P_R)
\right]
{\mu}
\nonumber \\
&&\times
\sum_{q=u,d} Q^q_{em} \bar{q} \gamma^{\alpha} {q}
\nonumber \\
&&+\frac{g_Z^2}{m_Z^2}
\sum_{q=u,d} \frac{Z_L^q +Z_R^q}{2} \bar{q} \gamma_{\alpha}q
{}~\bar{e} \gamma^{\alpha}(F_L P_L + F_R P_R) {\mu},
\end{eqnarray}
where the first term comes from the Penguin-type diagrams
of photon exchange and the second one Z boson exchange.
The coefficient $Q^q_{em}$ denotes the electric charge of the quark
$q$ and $Z_{L(R)}^q$ is Z boson coupling to the quark $q_{L(R)}$
such as Eq.~(\ref{Z_coupling}).

The box-type diagrams give
\begin{eqnarray}
{\cal L}^{box}_{eff}&=& e^2
\sum_{q=u,d}
\bar{q} \gamma_\alpha q
{}~~\bar{e} \gamma^\alpha
(
D_q^{L} P_L + D_q^{R} P_R
)
\mu,
\end{eqnarray}
with
\begin{eqnarray}
D_q^{L,R}=D_q^{(n)L,R}+D_q^{(c)L,R}~~~~~(q=u,d).
\end{eqnarray}
The coefficients are calculated to be
\begin{eqnarray}
e^2 D_q^{(n)L}&=&
\frac{1}{8}J_{4(A,B,X,Y)}
(
N_{\mu AX}^{R(l)*} N_{e BX}^{R(l)} N_{q AY}^{R(q)} N_{q BY}^{R(q)*}
-N_{\mu AX}^{R(l)*} N_{e BX}^{R(l)} N_{q AY}^{L(q)*} N_{q BY}^{L(q)}
)
\nonumber \\
&&-\frac{1}{4} M_{\tilde{\chi}_A^0} M_{\tilde{\chi}_B^0}I_{4(A,B,X,Y)}
(N_{\mu AX}^{R(l)*} N_{eBX}^{R(l)}
N_{qAY}^{L(q)} N_{qBY}^{L(q)*}
\nonumber \\
&&-N_{\mu AX}^{R(l)*} N_{eBX}^{R(l)}
N_{qAY}^{R(q)*} N_{qBY}^{R(q)}),
\\
D_q^{(n)R}&=&D_q^{(n)L}|_{L \leftrightarrow R}~~~~~(q=u,d),
\end{eqnarray}
and
\begin{eqnarray}
e^2 D_d^{(c)L}&=&
\frac{1}{8}J_{4(A,B,X,Y)} C_{\mu AX}^{R(l)*} C_{eBX}^{R(l)}
C_{dAY}^{R(d)} C_{dBY}^{R(d)*}
\nonumber \\
&&-\frac{1}{4} M_{\tilde{\chi}_A^-} M_{\tilde{\chi}_B^-}I_{4(A,B,X,Y)}
C_{\mu AX}^{R(l)*} C_{eBX}^{R(l)}
C_{dAY}^{L(d)} C_{dBY}^{L(d)*},
\\
e^2 D_u^{(c)L}&=&
-\frac{1}{8}J_{4(A,B,X,Y)} C_{\mu AX}^{R(l)*} C_{eBX}^{R(l)}
C_{uAY}^{L(u)*} C_{uBY}^{L(u)}
\nonumber \\
&&+\frac{1}{4} M_{\tilde{\chi}_A^-} M_{\tilde{\chi}_B^-}I_{4(A,B,X,Y)}
C_{\mu AX}^{R(l)*} C_{eBX}^{R(l)}
C_{uAY}^{R(u)*} C_{uBY}^{R(u)}.
\end{eqnarray}

Note that we only take account of the vector contributions for the
quark currents. The reason is given as follows. In the limit of the
low momentum transfer which is appropriate for the present case
$(q^2\simeq -m_\mu^2)$, we can treat the hadronic current in the
non-relativistic limit.  Furthermore, the contributions from the
coherent process dominates over the incoherent ones if we concentrate
on the relevant process such as $\mu +^{48}_{22}{\rm Ti}\rightarrow
e+^{48}_{22}{\rm Ti}$.  Then, the matrix element for the $\mu$-$e$
conversion process is dominated by the contribution from the vector
currents.

\subsection{Decay rates and conversion rate}
Now it is straightforward to calculate the decay rates and the
conversion rate, using the amplitudes (or the effective Lagrangian)
given in the above subsection.
\subsubsection{
$l_j^- \rightarrow l_i^-~\gamma$ decay rate}
 The decay rate for $l_j^- \rightarrow l_i^-~\gamma$
is easily calculated using the amplitude (\ref{Penguin}),
\begin{eqnarray}
\Gamma(l_j^- \rightarrow l_i^-~\gamma)
= \frac{e^2}{16 \pi} m_{l_j}^5 (|A_2^L|^2+|A_2^R|^2).
\end{eqnarray}

\subsubsection{
$l_j^- \rightarrow l_i^-~l_i^-~l_i^+$ decay rate}
Using the expressions for the amplitude,  we can calculate the
decay rate,
\begin{eqnarray}
\Gamma({l_j}^- {\rightarrow} l_i^-~l_i^-~l_i^+)
&=&\frac{e^4}{512 \pi^3}m_{l_j}^5
\left [
|A_1^L|^2+|A_1^R|^2-2(A_1^L A_2^{R*}+A_2^L A_1^{R*}+h.c.)
\right.
\nonumber \\
&&+(|A_2^L|^2+|A_2^R|^2)
\left(
\frac{16}{3}\ln{\frac{m_{l_j}}{2m_{l_i}}}-\frac{14}{9}
\right)
\nonumber \\
&&+\frac{1}{6}(|B_1^L|^2+|B_1^R|^2)
+\frac{1}{3}(|B_2^L|^2+|B_2^R|^2)
+\frac{1}{24}(|B_3^L|^2+|B_3^R|^2)
\nonumber \\
&&+6 (|B_4^L|^2+|B_4^R|^2)
-\frac{1}{2}(B_3^L B_4^{L*} + B_3^R B_4^{R*} + h.c.)
\nonumber \\
&&+\frac{1}{3}
(A_1^L B_1^{L*}+ A_1^R B_1^{R*}+A_1^L B_2^{L*}+A_1^R B_2^{R*} +h.c.)
\nonumber \\
&&
-\frac{2}{3}(A_2^R B_1^{L*}+A_2^L B_1^{R*} +A_2^L B_2^{R*}
+A_2^R B_2^{L*}+h.c.)
\nonumber \\
&&
+\frac{1}{3} \left \{
2 ( \left|F_{LL} \right|^2 +\left|F_{RR} \right|^2 )
+\left|F_{LR} \right|^2 + \left|F_{RL} \right|^2
\right.
\nonumber \\
&&+( B_1^L F_{LL}^*+B_1^R F_{RR}^*
+ B_2^L F_{LR}^* + B_2^R F_{RL}^* + h.c. )
\nonumber \\
&&+2 (
A_1^L F_{LL}^* +A_1^R F_{RR}^* +h.c.)
+( A_1^L F_{LR}^* +A_1^R F_{RL}^* +h.c. )
\nonumber \\
&&
\left. \left.
-4 ( A_2^R F_{LL}^* + A_2^L F_{RR}^* +h.c. )
-2 ( A_2^L F_{RL}^* + A_2^R F_{LR}^* +h.c. )
\right \}
\right ],
\nonumber \\
\label{amp-mu3e}
\end{eqnarray}
where
\begin{eqnarray}
F_{LL}&=&\frac{F_L Z_L^l}{m_Z^2 \sin^2 \theta_W \cos^2 \theta_W},
\\
F_{RR}&=& \left. F_{LL} \right|_{L\leftrightarrow R},
\\
F_{LR}&=&\frac{F_L Z_R^l}{m_Z^2 \sin^2 \theta_W \cos^2 \theta_W},
\\
F_{RL}&=& \left. F_{LR} \right|_{L\leftrightarrow R}.
\end{eqnarray}

Numerically, we find that a Penguin-type contribution involving
$A_2^L$ and $A_2^R$ dominates over the other contributions. In the
large $\tan\beta$ region, its effect is enhanced due to the same
mechanism as in the case of $l_j^- \rightarrow l_i^-~\gamma$ process.
Furthermore, even in the case where $\tan\beta$ is not so large,
the contribution of the Penguin-type diagram dominates over the box
contribution, because of the logarithmic term in Eq.~(\ref{amp-mu3e})
which is quite larger than the other terms.\footnote
{This logarithmic term is obtained as a result of the phase space
integration of the fermions in the final state, since we have an
infrared singularity in the limit of $m_{l_i}\rightarrow 0$.}
Then, the above formula is greatly simplified, and one finds a simple
relation
\begin{equation}
\frac{Br( l_j^- \rightarrow l_i^- l_i^- l_i^+)}
    {Br(l_j^- \rightarrow l_i \gamma)}
 \simeq \frac{\alpha}{8 \pi} \left(\frac{16}{3} \ln \frac{m_{l_j}}{2 m_{l_i}}
 -\frac{14}{9} \right).
\label{penguinphase}
\end{equation}

\subsubsection
{$\mu$-$e$ conversion rate ($\mu +(A,Z) \rightarrow e + (A,Z)$)}

Once we know the effective Lagrangian relevant to this process at the
quark level, we can calculate the conversion rate~\cite{mu->e},
\begin{eqnarray}
\nonumber
\Gamma(\mu\rightarrow e)
&=& 4\alpha^5 \frac{Z_{eff}^4}{Z}|F(q)|^2~m_{\mu}^5
\left [
|Z(A_1^L-A_2^R)-(2Z+N){\bar D}_u^L-(Z+2N){\bar D}_d^L|^2
\right.
\nonumber \\
&&\left.+|Z(A_1^R-A_2^L)-(2Z+N){\bar D}_u^R-(Z+2N){\bar D}_d^R|^2
\right ],
\end{eqnarray}
where
\begin{eqnarray}
{\bar D}_q^L&=&D_q^L + \frac{Z_L^q+Z_R^q}{2}
\frac{F_L}{m_Z^2 \sin^2 \theta_W \cos^2 \theta_W},
\\
{\bar D}_q^R&=&D_q^L |_{L \leftrightarrow R}~~~~~~(q=u,d),
\end{eqnarray}
and $Z$ and $N$ denote the proton and neutron numbers in a nucleus,
respectively. $Z_{eff}$ has been determined in~\cite{z-eff} and $F(q^2)$
is the nuclear form factor. In $^{48}_{22}{\rm{Ti}}, Z_{eff}=17.6,
F(q^2\simeq -m_{\mu}^2) \simeq 0.54$ \cite{mu->e}.

\section{Results of the Numerical Calculations}

In this section, we present results of our numerical analysis.

As was discussed in Section 2, we assume the universal scalar masses.
Also for simplicity, we consider the so-called GUT relation among the
gaugino masses
\begin{eqnarray}
\frac{M_1}{g_1^2}=\frac{M_2}{g_2^2}=\frac{M_3}{g_3^2}.
\end{eqnarray}
Then the SUSY breaking terms have four free parameters; the universal
scalar mass ($m_0$), the $SU(2)_L$ gaugino mass at low energies
($M_2$), the universal $A$-parameter ($A=a m_0$) and mixing parameter
of the two Higgs bosons ($B$).

Concerning the SUSY invariant Higgs mass $\mu$ and $B$-parameter which
parameterize the mixing among $h_1$ and $h_2$, we determined them so
that the two Higgs doublets have correct vacuum expectation values
$\langle h_1 \rangle =v\cos\beta/\sqrt{2}$ and $\langle h_2 \rangle
=v\sin\beta/\sqrt{2}$. With this radiative electroweak symmetry breaking
condition~\cite{PTP68-927}, we determine the mass spectra and mixing
matrices of the superparticles. Then, we carefully investigate the
parameter space where $\tan\beta$ is large and masses of superparticles
(especially, sleptons and electroweak gauginos) are quite light enough
to enhance the LFV rates. As a result, we found that there indeed exists
parameter space where the above conditions are satisfied.  We checked,
for $M_2=80$GeV, $\tan
\beta$ can be as large as about 50.\footnote
{Throughout this paper, we take the top quark mass $m_t=174$ GeV~\cite{PDG}.
Also we take the bottom quark mass $m_b=4.25$GeV~\cite{PRepC87-77}, which
corresponds to 3.1 GeV at the $Z$ mass scale.}
This result implies that there
are regions in the parameter space where the LFV processes have large
branching ratios due to the large $\tan \beta$ enhancement
mechanism.\footnote
{Note that the situation here contrasts to the case of the Yukawa
unification where the radiative breaking with the universal scalar
mass requires heavy superparticle spectrum, larger than, say, 500 GeV
\cite{Nakano}.}

We also put constraints from experiments.  Besides our requirement that
the lightest superparticle be neutral, we use consequences of the
negative searches for the superparticles~\cite{PDG}. We also impose a
constraint on SUSY contribution to the anomalous magnetic dipole-moment
of the muon~\cite{g-2,hep-ph/9507386}.  The experimental value of
$\frac{1}{2}(g-2)$ is $1165923 (8.4)\times 10^{-9}$~\cite{PDG}.  On the
other hand, the theoretical prediction of standard model is
$11659180(15.3)\times 10^{-10}$ or
$11659183(7.6)\times10^{-10}$~\cite{hep-ph/9507386}, where the
difference is due to different estimates of hadronic contributions. In
our paper, we adopt the first one in order to derive conservative bound.
Therefore, the SUSY contribution should be constrained as
\begin{eqnarray}
-26.7 \times 10^{-9} < (g-2)_{\mu}^{SUSY} <46.7 \times 10^{-9},
\end{eqnarray}
where two sigma experimental error is considered.  The SUSY
contribution is shown in Fig.~7.  Here, we take the parameter $a=0$ at
the gravitational scale and $M_2=100$ GeV at low energies. The
horizontal line is taken to be the left-handed selectron mass with the
D-term contribution, which we denote by $m_{\tilde{e}_L}$. One finds
that a significant region of the parameter space is excluded by this
constraint in the large $\tan \beta$ region. This is because the same
enhancement mechanism as the LFV processes works in the diagrams
contributing to the $g-2$. For completeness, we will give formulas of
the contribution of the superparticle loops to the anomalous magnetic
dipole-moment in Appendix C.

Let us now discuss the branching ratios for each LFV process.
First we consider the case where the neutrino mixing matrix is
described by the KM matrix.

\subsection{ {\em  Case 1\/}) $V= V_{KM}$}

 As the first trial, we shall consider the case where $V=V_{KM}$,
where we take $s_{12}=0.22$, $s_{23}=0.04$ and $s_{13}=0.0035$ in the
standard notation~\cite{PDG}.  We ignore the possible
Kobayashi-Maskawa complex phase and consider $V$ to be real, for
simplicity.  The eigenvalues of the neutrino Yukawa couplings are
assumed to be equal to those of the up-type quarks at the
gravitational scale.  Since the magnitude of the top quark Yukawa coupling
is close to its perturbative bound, this ansatz will maximize the
magnitude of LFV in the slepton mass matrix.  Also, to determine
the right-handed neutrino Majorana mass $M_R$, we fix the tau neutrino
mass at 10 eV so that it constitutes the hot component of the dark
matter of the Universe.  In this case, $M_R$ is about $10^{12-13}$ GeV.

Solving the RGEs numerically, we obtain the mass squared matrix for
the $SU(2)_L$ doublet sleptons at the electroweak scale
\begin{equation}
m_{\tilde L}^2 \simeq
\left(
\begin{array}{ccc}
   1.00  & (0.30-0.43) \times 10^{-4} & -(0.74-1.07) \times 10^{-3}
\\
  (0.30-0.43) \times 10^{-4} & 1.00 & -(0.54-0.78) \times 10^{-2}
\\
-(0.74-1.07) \times 10^{-3} & -(0.54-0.78) \times 10^{-2} & 0.77-0.80
\end{array} \right) \times m_0^2,
\label{mL-KM}
\end{equation}
where $\tan \beta$ varies from 3 to 30, $M_2=0$ and $a=0$.  For a
non-vanishing $M_2$, the diagonal elements of the above matrix become
larger and the flavor-violating off-diagonal elements become
relatively less important, as the gaugino mass gets larger. Effect of
non-vanishing $a$-parameter can be seen from
Eq.~(\ref{slepton-approx}), which does not change the result
drastically. In the following numerical calculations we will take
$a=0$.

We find in Eq.~(\ref{mL-KM}) the off-diagonal elements in the mass
matrix are small. This is because the off-diagonal slepton masses are
proportional to $V^*_{3i} V^{3j}$ in the case of hierarchical neutrino
masses, which are small if we assume that $V$ is equal to the KM
matrix.  Nevertheless, as will be shown shortly, the enhancement in
the large $\tan \beta$ region yields large branching ratios for the
LFV processes, which is close to the present experimental upper
bounds.

\subsubsection{ $\mu \rightarrow e \gamma$}

Result of our computation on the branching ratio $Br (\mu\rightarrow
e\gamma)$ is shown in Fig.~8 for $M_2=100$ GeV. The horizontal line is
taken to be the left-handed selectron mass with the D-term
contribution, $m_{\tilde{e}_L}$. Real lines are for $\mu >0$, while
dashed lines for $\mu <0$.  We can find that the branching ratios are
rather insensitive to the choice of the sign of the $\mu$-parameter,
in particular when $\tan\beta$ is large.  For the large $\tan\beta$
case, some regions of small slepton masses are excluded by the
constraint from $g-2$.  As can be seen from Fig.~7, it is less
stringent for $\mu >0$ case than $\mu <0$ case.\footnote
{Here, we should comment that the SUSY contribution to the
$b\rightarrow s\gamma$ process is also significant and some part of
the parameter space should be excluded
\cite{hep-ph/9507386,b-sgamma,PLB300-300}.
However, it is complicated to estimate the SUSY contribution to the
$b\rightarrow s\gamma$ process, since the chargino loop can contribute
either constructively or destructively to the others, especially
charged Higgs boson loop. Thus, it seems to us that to determine which
regions of the parameter space are really eliminated contains some
delicate issues as discussed by Ref.~\cite{PLB300-300}. We believe
that such an analysis is out of the scope of our paper, but a work in
a future communication. Thus, we do not use the constraint from the
$b\rightarrow s\gamma$ process.}
One can see that even if we impose this constraint, the branching
ratio can be as large as $ 10^{-11}$, which is very close to the
present experimental bound $Br(\mu\rightarrow e \gamma)|_{\rm exp} <
4.9\times 10^{-11}$.  For smaller value of $\tan\beta$, the branching
ratio reduces obeying $\propto\tan^2\beta$.

We compared the chargino loop contribution with the neutralino loop
contribution and found that the former dominates.  This is important
when we compare our results with the case of $SU(5)$ grand unification.
(See Section 5.)


In Fig.~9, we show the case of $M_2=200$ GeV.  The maximum of the
branching ratio is about $10^{-12}$ for $\tan \beta=30$, about one
order of magnitude smaller than the $M_2=100 $ GeV case.  We also
studied the case $M_2=80$ GeV, and found that the branching ratio is
about factor 2 larger than the $M_2=100 $ GeV case.

\subsubsection{ $\mu^- \rightarrow e^- e^- e^+$}

Next, let us consider the process $\mu^- \rightarrow e^- e^- e^+$.
Currently the experimental upper bound on the branching ratio of this
process is $1.0 \times 10^{-12}$~\cite{PDG}. We show results of our
calculation to this process in Fig.~10 for $M_2=100$ GeV.  The
branching ratio has the maximum of $\sim 10^{-13}$ for the large $\tan
\beta$ with the small gaugino mass.  One can check that this process
is dominated by the Penguin-type diagrams. Indeed compared with the
branching ratio of $\mu
\rightarrow e \gamma$, one finds a simple relation
\begin{equation}
  \frac{Br(\mu \rightarrow 3e)}{Br(\mu \rightarrow e \gamma)}
  \sim  7 \times 10^{-3},
\end{equation}
which is in  agreement with the ratio expected by the dominance of the
Penguin-type diagrams, Eq.~(\ref{penguinphase}).

\subsubsection{ $\mu$-$e$ conversion in $^{48}_{22}$Ti}

Experimentally, $\mu$-$e$ conversion rate in nuclei is also
constrained strongly. The experimental upper bound on the conversion
rate with the target $^{48}_{22}$Ti reaches 4.3$\times
10^{-12}$~\cite{PDG}.  We show results of our calculation to this
process in Fig.~11 for $M_2=100$ GeV. The branching ratio takes its
maximal value of $\sim 10^{-13}$ in the parameter region where
$\tan\beta$ is large and the gaugino masses are small.  On the other
hand, for the small $\tan
\beta$ and $\mu<0$ the cancelation among the diagrams occurs and the
event rate damps rapidly. The Penguin-type diagram is not dominant in the
small $\tan\beta$ region because there is not the same logarithmic
enhancement as $\mu^-
\rightarrow e^- e^- e^+$.

\subsubsection{ $\tau \rightarrow \mu \gamma$ }

Finally we would present our result for $\tau \rightarrow \mu \gamma$
in Fig.~12.  We find with $M_2=100$ GeV, the branching ratio is as
large as $10^{-7}$, one and a half order of magnitude smaller than the
present experimental bound $Br(\tau \rightarrow \mu \gamma)|_{\rm exp}
<4.2\times 10^{-6}$~\cite{PDG}.  Similar to the case of $\mu
\rightarrow e \gamma$, it can be seen that the branching ratio is
proportional to $\tan
\beta$ squared.

\subsection{ {\em  Case 2\/}) Neutrino mixing implied by atmospheric
neutrino deficit}

A class of solutions to the atmospheric and solar neutrino deficits
requires a maximal mixing of the tau and muon neutrinos, yielding a
large off-diagonal element in the slepton mass matrix.  The neutrino
mixing matrix we take in this example is
\begin{equation}
  V\simeq \left(
   \begin{array}{ccc}
     1.00 & 0.87 \times 10^{-1} & - \\
     -0.66 \times 10^{-1} & 0.755 & 0.656 \\
      - & -0.656 & 0.755
   \end{array} \right)
\label{V-atmospheric}
\end{equation}
and the tau neutrino mass is assumed to be 0.4 eV \cite{pakvasa}.
Here, we only consider the generation mixing of the second and third
generations and ignore the others.  The (1,3) and (3,1) elements of
the mixing matrix cannot be determined from the solar and atmospheric
neutrino deficits. This uncertainty, however, does not matter if we
only consider the LFV process among the second and third generations.
As in the {\em case 1}), we assume the magnitude of the third
generation neutrino Yukawa coupling $f_{\nu 3}$ is equal to the top
quark Yukawa coupling at the gravitational scale.  The latter choice
will give us a maximum violation of LFV in the slepton mass matrix.

The result for $Br (\tau \rightarrow \mu \gamma)$ is shown in Fig.~13.
We find that in some portion of the parameter space, the branching
ratio exceeds the present experimental upper bound, in particular when
$\tan\beta$ is large and the superparticles are light.

\section{Conclusions and Discussion}

In this paper, we have considered LFV in the minimal supersymmetric
standard model (MSSM) with the right-handed neutrino multiplets. In
the presence of the Yukawa couplings of the right-handed neutrinos,
the left-handed slepton mass matrix, $m_{\tilde L}^2$, loses its
universal property even if we assume the minimal supergravity type
boundary condition on sfermion masses.  In our case, due to the
renormalization effect, as can be seen from Eq.~(\ref{mL-KM}), we
obtain LFV in $m_{\tilde L}^2$ as well as smaller value of (3,3)
element of $m_{\tilde L}^2$ compared with the other diagonal elements,
which is typical feature of the case with right-handed
neutrino~\cite{PLB321-56}.  We have calculated the interaction rates
for the various LFV processes with the full diagonalization of the
slepton mass matrices and of the chargino and neutralino mass
matrices.  We emphasized the enhancement of the interaction rates for
large $\tan \beta$, the ratio of the VEVs of the two Higgs doublets.
This enhancement is originated to the fact that there is a freedom to
pick up one of two vacuum expectation values in the MSSM in the
magnetic dipole-moment type diagrams.  For example, for the process
$l_j \rightarrow l_i \gamma$, the diagrams of the type Fig.~1(c) and
Fig.~2 give the enhancement.  Even when the mixing matrix in the
lepton sector has a similar structure as the KM-matrix of the quark
sector, the enhancement mechanism can make the branching ratios close
to the present experimental bounds.

It is interesting to compare the LFV processes induced by the
right-handed neutrino Yukawa couplings with those in the minimal
$SU(5)$ grand unified theory~\cite{PLB338-212,NPB445-219}. In the
latter case, the renormalization-group flow above the GUT scale
results in LFV in the $SU(2)_L$ singlet (right-handed) slepton masses.
Let us consider, for example, the resulting branching ratio of
$\mu\rightarrow e\gamma$.  The diagrams which will give the
enhancement in the large $\tan \beta$ region are similar to Fig.~1(c)
and 2(a). The important difference from the previous case is on
Fig.~2.  Now, only the diagrams involving the bino contributes, since
the wino does not couple to the singlet sleptons. In this case, we can
see that contributions coming from the two diagrams Fig.~1(c) and
Fig.~2(a) have opposite signs, and thus partially cancel out with each
other.  Numerical result is shown in Fig.~14.  The horizontal line is
the mass of the right-handed selectron with the D-term contribution,
$m_{\tilde{e}_R}$.  Here, we have taken $M_2=100$ GeV.  The branching
ratio never exceeds $10^{-13}$, more than two orders of magnitude
beneath the present experimental upper bound.  Also one finds regions
where the branching ratio becomes very small due to the cancelation
explained above.

What happens if the standard model with the right-handed neutrinos is
embedded in the framework of $SU(5)$ GUT?  In this case, both the mass
matrix of the left-handed sleptons and that of the right-handed ones
have LFV.  The situation is quite similar to the case of $SO(10)$
GUT~\cite{NPB445-219,LFV_SO10}.  For example, if we consider the
$\mu\rightarrow e \gamma$, the dominant diagram will be similar to
Fig.~1(c), which however picks up $(m_{LR}^2)^3_3$, proportional to
tau-lepton mass.  Thus we expect further enhancement in the branching
ratio by $(m_{\tau}/m_{\mu})^2$ compared to the case we studied in
this paper.

To conclude our paper, we should emphasize that the branching ratios
of the LFV processes induced by the right-handed neutrino Yukawa
couplings can be close to the present experimental bounds and can be
within the reach of future experiments.  Efforts of searching for
these  LFV signals should be encouraged.

\section*{Acknowledgment}
We would like to thank T.~Yanagida for useful discussions. We are also
grateful to T.~Goto and P.~Nath for discussion on the $b\rightarrow
s\gamma$ process, and to S.~Orito for a comment on the anomalous
magnetic dipole-moment of the muon. One of authors (J.H.) is a fellow
of the Japan Society for the Promotion of Science.

\appendix

\section{Renormalization Group Equations}
In this appendix, we give the one-loop renormalization group equations
(RGEs) for the Yukawa couplings and the soft SUSY breaking terms in
the scalar potential.  The RGEs for the gauge coupling constants and
the gaugino masses are unchanged at the one-loop level, since the
right-handed neutrinos are singlet under the standard-model gauge symmetry.
\begin{itemize}
\item{Yukawa coupling constants}
\begin{eqnarray}
\mu \frac{d}{d \mu} f_l^{ij}&=& \frac{1}{16 \pi^2}
\left[
\left \{
-\frac{9}{5}g_1^2 -3 g_2^2 + 3 Tr(f_d f_d^{\dagger})
+ Tr(f_l f_l^{\dagger})
\right \} f_l^{ij} \right.
\nonumber \\
&&\left.
+3(f_l f_l^{\dagger} f_l)^{ij} + (f_l f_{\nu}^{\dagger} f_{\nu})^{ij}
\right],
\\
\mu \frac{d}{d \mu} f_{\nu}^{ij}&=& \frac{1}{16 \pi^2}
\left[
\left \{
-\frac{3}{5} g_1^2 -3 g_2^2 +3 Tr(f_{u} f_{u}^{\dagger})
+Tr(f_{\nu} f_{\nu}^{\dagger})
\right \} f_{\nu}^{ij}
\right.
\nonumber \\
&&\left.
+3(f_{\nu} f_{\nu}^{\dagger} f_{\nu})^{ij}
+ (f_{\nu}f_{l}^{\dagger} f_{l})^{ij}
\right ].
\end{eqnarray}
\item{Soft breaking terms}
\begin{eqnarray}
\mu \frac{d}{d \mu} (m^2_{\tilde L})_i^j&=&
\frac{1}{16 \pi^2} \left [
\left(m^2_{\tilde L} f_l^{\dagger} f_l
+f_l^{\dagger} f_l m^2_{\tilde L} \right)_i^j
+\left(m^2_{\tilde L} f_{\nu}^{\dagger} f_{\nu}
+f_{\nu}^{\dagger} f_{\nu} m^2_{\tilde L} \right)_i^j
\right.
\nonumber \\
&&+2 \left( f_l^{\dagger} m^2_{\tilde e} f_l
+{\tilde m}^2_{h1}f_l^{\dagger}f_l
+A_l^{\dagger} A_l \right)_i^j
\nonumber \\
&&+2 \left( f_{\nu}^{\dagger} m^2_{\tilde \nu} f_{\nu}
+{\tilde m}^2_{h2}f_{\nu}^{\dagger} f_{\nu}
+A_{\nu}^{\dagger} A_{\nu} \right)_i^j
\nonumber \\
&&
\left.
-\left (\frac{6}{5}g_1^2 \left| M_1 \right|^2
+6 g_2^2 \left| M_2 \right|^2 \right) \delta_i^j
-\frac{3}{5} g_1^2 S \delta_i^j
\right ],
\\
\mu \frac{d}{d \mu} (m^2_{\tilde e})^i_j&=&
\frac{1}{16 \pi^2} \left [
2 \left(m^2_{\tilde e} f_l f_l^{\dagger}
+f_l f_l^{\dagger} m^2_{\tilde e} \right)^i_j
\right.
\nonumber \\
&&
\left.
+4 \left( f_l m^2_{\tilde L} f_l^{\dagger}
+{\tilde m}^2_{h1}f_l f_l^{\dagger}
+A_l A_l^{\dagger}\right)^i_j
\right.
\nonumber \\
&&\left.
-\frac{24}{5} g_1^2 \left| M_1 \right|^2 \delta^i_j
+\frac{6}{5}g_1^2 S \delta^i_j
\right],
\nonumber \\
\mu \frac{d}{d \mu} (m^2_{\tilde \nu})^i_j&=&
\frac{1}{16 \pi^2} \left [
2 \left(m^2_{\tilde \nu} f_{\nu} f_{\nu}^{\dagger}
+f_{\nu}f_{\nu}^{\dagger} m^2_{\tilde \nu} \right)^i_j
\right.
\nonumber \\
&&
\left.
+4 \left( f_{\nu} m^2_{\tilde L} f_{\nu}^{\dagger}
+{\tilde m}^2_{h2}f_{\nu} f_{\nu}^{\dagger}
+A_{\nu} A_{\nu}^{\dagger}\right)^i_j
\right],
\\
\mu \frac{d}{d \mu} A_l^{ij}&=&
\frac{1}{16 \pi^2} \left[ \left\{
-\frac{9}{5} g_1^2 -3 g_2^2+ 3 Tr(f_d^{\dagger} f_d)
+Tr(f_l^{\dagger} f_l) \right \} A_l^{ij}
\right.
\nonumber \\
&&+2 \left\{
-\frac{9}{5} g_1^2 M_1 -3 g_2^2 M_2 + 3 Tr(f_d^{\dagger} A_d)
+Tr(f_l^{\dagger} A_l) \right \} f_l^{ij}
\nonumber \\
&&\left.
+4 (f_l f_l^{\dagger} A_l)^{ij} + 5 (A_l f_l^{\dagger} f_l)^{ij}
+2(f_l f_{\nu}^{\dagger} A_{\nu})^{ij} +
 (A_l f_{\nu}^{\dagger} f_{\nu})^{ij}
\right],
\\
\mu \frac{d}{d \mu} A_{\nu}^{ij}&=&
\frac{1}{16 \pi^2} \left[ \left\{
-\frac{3}{5}g_1^2 -3g_2^2 +3 Tr(f_u^{\dagger} f_u)
+Tr(f_{\nu}^{\dagger} f_{\nu}) \right \} A_{\nu}^{ij}
\right.
\nonumber \\
&&+2 \left\{
-\frac{3}{5} g_1^2 M_1 -3 g_2^2 M_2 + 3 Tr(f_u^{\dagger} A_u)
+Tr(f_{\nu}^{\dagger} A_{\nu}) \right \} f_{\nu}^{ij}
\nonumber \\
&&\left.
+4(f_{\nu} f_{\nu}^{\dagger}A_{\nu})^{ij}
+5(A_{\nu}f_{\nu}^{\dagger} f_{\nu})^{ij}
+2(f_{\nu} f_l^{\dagger}A_l)^{ij}
+(A_{\nu} f_l^{\dagger} f_l)^{ij}
\right],
\end{eqnarray}
where
\begin{eqnarray}
S=Tr(m_{\tilde Q}^2 + m_{\tilde d}^2 -2 m_{\tilde u}
-m_{\tilde L}^2 +m_{\tilde e}^2)
- {\tilde m}^2_{h1}+{\tilde m}^2_{h2}.
\end{eqnarray}
\end{itemize}
Here, we followed the GUT convention for the normalization of $U(1)_Y$
gauge coupling constant $g_1$, such as $g_Y^2=\frac{3}{5}g_1^2$.

\section{Interaction of gaugino-sfermion-fermion}

In this appendix, we give our notations and conventions adopted in
Section 3 and give vertices relevant for our calculation.

Let us first discuss fermions.  We denote by $l_i$, $u_i$ and $d_i$
the fermion mass eigenstates with the obvious meaning.  The subscript
$i$ ($i=1,2,3)$ represents the generation.  As for the neutrinos,
their masses are small and negligible.  In our convention, $\nu_i$ is
the $SU(2)_L$ isodoublet partner to $e_{Li}$.

Next we consider sfermions. Let $\tilde f_{Li}$ and $\tilde f_{Ri}$ be the
superpartners of $f_{Li}$ and $f_{Ri}$, respectively.  Here, $f$ stands
for $l$, $u$ or $d$.  The mass matrix
for the sfermions can be written in the following form,
\begin{equation}
     \left ( \tilde f^{\dagger}_{L}, \tilde f^{\dagger}_R \right )
     \left ( \begin{array}{cc}
               m_L^2    & m_{LR}^{2 {\sf T}} \\
               m_{LR}^{2} & m_R^2
             \end{array}                  \right )
     \left (  \begin{array}{c}
              \tilde f_{L} \\ \tilde f_R
              \end{array}                 \right ),
\end{equation}
where $m_L^2$ and $m_R^2$ are $3 \times 3$ hermitian matrices and
$m_{LR}^2$ is a $3\times 3$  matrix. These elements are given from
Eqs.~(\ref{superpotential},\ref{softbreaking}) as following,
\begin{eqnarray}
m_L^2 &=&
m_{\tilde f_L}^2 + m_f^2
+ m_Z^2 \cos 2 \beta (T_{3L}^f -Q_{em}^f \sin^2 \theta_W),
\\
m_R^2 &=&
m_{\tilde f_R}^2
+ m_f^2 - m_Z^2 \cos 2 \beta (T_{3R}^f -Q_{em}^f \sin^2 \theta_W),
\\
m_{LR}^2 &=&
\left\{
\begin{array}{c}
-A_f v \sin\beta /\sqrt{2}  - m_f \mu \cot\beta ~~~~ (f=u),
\\
A_f v \cos\beta /\sqrt{2} - m_f \mu \tan\beta ~~~~ (f=d,l),
\end{array}
\right.
\end{eqnarray}
where $T_{3L(R)}^f$ and $Q^f_{em}$ are weak isospin and electric charge
respectively. Here, $m_{\tilde f_L}^2 = m_{\tilde Q}^2$ for squarks,
$m_{\tilde f_L}^2 = m_{\tilde L}^2$ for sleptons, and
$m_{\tilde f_R}^2$ are each right-handed sfermion soft-breaking
masses. We assume the above mass matrix to be real.
This is, in general, not diagonal and include mixing between
different generations.  We diagonalize the mass matrix
${\cal  M}^2$ by  a $6 \times
6$ real orthogonal matrix $U^f$ as
\begin{equation}
    U^f {\cal M}^2 U^{f {\sf T}} =({\rm diagonal}),
\end{equation}
and we denote its eigenvalues by $m^2_{\tilde f_X}$ ($X= 1, \cdots,
6$).  The mass eigenstate is then written as
\begin{equation}
   \tilde f_X = U^f_{X,i} \tilde f_{Li} + U^f_{X,i+3} \tilde f_{Ri},
   \hspace{1.5cm} (X=1, \cdots, 6).
\end{equation}
Conversely, we have
\begin{eqnarray}
   \tilde f_{Li} = &U^{f{\sf T}}_{iX} \tilde f_X & =U^{f}_{Xi} \tilde f_X,
\\
   \tilde f_{Ri} =& U^{f{\sf T}}_{i+3,X} \tilde f_X & = U^f_{X,i+3} \tilde
   f_X.
\end{eqnarray}

An attention should be paid to the  neutrinos since there is no
right-handed sneutrino in the MSSM.  Let $\tilde \nu_{Li}$ be the
superpartner of the neutrino $\nu_{i}$.  The mass eigenstate
$\tilde \nu_X$ ($X=1,2,3$) is related to  $\tilde \nu_{Li}$ as
\begin{equation}
   \tilde \nu_{Li}=U^{\nu}_{Xi} \tilde \nu_{X}.
\end{equation}

We now turn to charginos.  The mass matrix of the charginos is given by
\begin{equation}
  -{\cal L}_m =
     \left ( \overline{\tilde W^-_R}~ \overline{\tilde H^-_{2R}} \right )
     \left (  \begin{array}{cc}
                M_2            & \sqrt{2} m_W \cos \beta \\
                \sqrt{2}m_W \sin \beta &  \mu
              \end{array}                                 \right)
     \left (  \begin{array}{c}
              \tilde W^-_L    \\ \tilde H_{1L}^-
              \end{array}                                 \right) +h.c..
\end{equation}
This matrix $M_C$ is diagonalized by $2 \times 2$ real orthogonal matrices
$O_L$ and $O_R$ as
\begin{equation}
     O_R M_C O_L^{\sf T} =({\rm diagonal}).
\end{equation}
Define
\begin{equation}
   \left( \begin{array}{c}
           \tilde \chi^-_{1L} \\
           \tilde \chi^-_{2L}
           \end{array}                 \right)
  =O_L   \left( \begin{array}{c}
                \tilde W^-_L  \\
                \tilde H^-_{1L}
                \end{array}            \right),
\hspace{1.5cm}
    \left( \begin{array}{c}
           \tilde \chi^-_{1R} \\
           \tilde \chi^-_{2R}
           \end{array}                 \right)
  =O_R   \left( \begin{array}{c}
                \tilde W^-_R  \\
                \tilde H^-_{2R}
                \end{array}            \right).
\end{equation}
Then
\begin{equation}
    \tilde \chi^-_A  =\tilde \chi^-_{AL} + \tilde \chi^-_{AR}
\hspace{1.5cm}
    (A=1,2)
\end{equation}
forms a Dirac fermion with mass $M_{\tilde \chi^-_A}$.

Finally we consider neutralinos.  The mass matrix of the neutralino
sector is given by
\begin{equation}
 -{\cal L}_m =
  \frac{1}{2}
  \left( \tilde B_L \tilde W^0_L \tilde H^0_{1L} \tilde H^0_{2L}
  \right)
   M_N
   \left(   \begin{array}{c}
            \tilde B_L  \\
            \tilde W^0_L \\
            \tilde H^0_{1L} \\
            \tilde H^0_{2L}
            \end{array}                  \right)  +h.c.,
\end{equation}
where
\begin{equation}
   M_N=
   \left(
   \begin{array}{cccc}
     M_1    & 0 & -m_Z\sin\theta_W\cos\beta & m_Z\sin\theta_W\sin\beta \\
     0 & M_2 & m_Z\cos\theta_W\cos\beta & -m_Z\cos\theta_W\sin\beta \\
     -m_Z\sin\theta_W\cos\beta & m_Z\cos\theta_W\cos\beta & 0 & -\mu
     \\
     m_Z\sin\theta_W\sin\beta & -m_Z\cos\theta_W\sin\beta & -\mu & 0
   \end{array}            \right).
\end{equation}
The diagonalization is done by a real orthogonal matrix $O_N$,
\begin{equation}
    O_N M_N O_N^{\sf T} = {\rm diagonal}.
\end{equation}
The mass eigenstates are given by
\begin{equation}
         \tilde \chi^0_{AL} =(O_N)_{AB} \tilde X^0_{BL}
\hspace{1.5cm}
           (A,B=1, \cdots ,4)
\end{equation}
where
\begin{equation}
   \tilde X^0_{AL} = ( \tilde B_L, \tilde W^0_L, \tilde H^0_{1L},
   \tilde H^0_{2L}).
\end{equation}
We have thus Majorana spinors
\begin{equation}
   \tilde \chi^0_A = \tilde \chi^0_{AL} + \tilde \chi^0_{AR},
{}~~~~(A=1, \cdots ,4)
\end{equation}
 with mass $M_{\tilde \chi^0_A}$.

We now give the interaction Lagrangian of fermion-sfermion-chargino,
\begin{eqnarray}
   {\cal L}_{\rm int}& = &
     \bar l_i (C^{R (l)}_{iAX} P_R+ C^{L(l)}_{iAX} P_L )
     \tilde \chi^-_A \tilde \nu_X
\nonumber \\
  & &+ \bar \nu_i (C^{R (\nu)}_{iAX} P_R+ C^{L(\nu)}_{iAX} P_L )
     \tilde \chi^+_A \tilde l_X
\nonumber \\
  & &+ \bar d_i (C^{R (d)}_{iAX} P_R+ C^{L(d)}_{iAX} P_L )
     \tilde \chi^-_A \tilde u_X
\nonumber \\
  & &+ \bar u_i (C^{R (u)}_{iAX} P_R+ C^{L(u)}_{iAX} P_L )
     \tilde \chi^+_A \tilde d_X
   +h.c.,
\end{eqnarray}
where the coefficients are
\begin{eqnarray}
 C^{R(l)}_{iAX}& =& -g_2(O_R)_{A1} U^{\nu}_{X,i},
\nonumber \\
 C^{L(l)}_{iAX}& = & g_2\frac{m_{l_i}}{\sqrt{2}m_W\cos\beta}(O_L)_{A2}
                    U^{\nu}_{X,i},
\nonumber \\
 C^{R(\nu)}_{iAX} &= & -g_2(O_L)_{A1} U^l_{X,i},
\nonumber \\
C^{L(\nu)}_{iAX} &= &g_2\frac{m_{l_i}}{\sqrt{2}m_W\cos\beta}(O_L)_{A2}
                     U^l_{X,i+3},
\nonumber \\
 C^{R(d)}_{iAX} & = & g_2 \{ -(O_R)_{A1} U^u_{Xi}
+\frac{m_{u_i}}{\sqrt{2}m_W\sin\beta} (O_R)_{A2} U^u_{X,i+3} \},
\nonumber \\
 C^{L(d)}_{iAX}& = & g_2 \frac{m_{d_i}}{\sqrt{2}m_W\cos\beta}(O_L)_{A2}
                    U^u_{X,i},
\nonumber \\
 C^{R(u)}_{iAX} & = & g_2 \{ -(O_L)_{A1}U^d_{X,i}
+\frac{m_{d_i}}{\sqrt{2}m_W\cos\beta}(O_L)_{A2} U^d_{X,i+3} \},
\nonumber \\
 C^{L(u)}_{iAX} & = & g_2 \frac{m_{u_i}}{\sqrt{2}m_W\sin \beta}
                    (O_R)_{A2} U^d_{X,i}.
\end{eqnarray}

The interaction Lagrangian of fermion-sfermion-neutralino is similarly
written as
\begin{equation}
  {\cal L}_{\rm int}
=  \bar f_i (N^{R(f)}_{iAX} P_R +N^{L(f)}_{iAX} P_L)
   \tilde \chi^0_A \tilde f_X
\end{equation}
where $f$ stands for $l,\nu,d$ and $u$.  The coefficients
 are
\begin{eqnarray}
  N^{R(l)}_{iAX}&=& -\frac{g_2}{\sqrt{2}} \{
       [-(O_N)_{A2} -(O_N)_{A1} \tan \theta_W] U^l_{X,i}
        + \frac{m_{l_i}}{m_W\cos\beta} (O_N)_{A3} U^l_{X,i+3} \},
\nonumber \\
  N^{L(l)}_{iAX} &=& -\frac{g_2}{\sqrt{2}} \{
           \frac{m_{l_i}}{m_W\cos\beta} (O_N)_{A3} U^{l}_{x,i}
           +2 (O_N)_{A1} \tan \theta_W U^{l} _{X,i+3} \},
\nonumber \\
  N^{R(\nu)}_{iAX} &=& -\frac{g_2}{\sqrt{2}}
             [(O_N)_{A2}-(O_N)_{A1} \tan\theta_W] U^{\nu}_{X,i},
\nonumber \\
  N^{L(\nu)}_{iAX}&=&0,
\nonumber \\
  N^{R(d)}_{iAX}&=& -\frac{g_2}{\sqrt{2}} \{
      [-(O_N)_{A2}+\frac{1}{3}(O_N)_{A1} \tan \theta_W] U^d_{X,i}
     +\frac{m_{d_{i}}}{m_W \cos\beta}(O_N)_{A3} U^d_{X,i+3} \},
\nonumber \\
  N^{L(d)}_{iAX}&=& -\frac{g_2}{\sqrt{2}}
     \{ \frac{m_{d_{i}}}{m_W\cos\beta}(O_N)_{A3} U^d_{X,i}
       +\frac{2}{3} \tan \theta_W (O_N)_{A1} U^d_{X,i+3} \},
\nonumber \\
  N^{R(u)}_{iAX}&=& -\frac{g_2}{\sqrt{2}}
         \{ [(O_N)_{A2} +\frac{1}{3} (O_N)_{A1} \tan \theta_W]
         U^u_{X,i}
        +\frac{m_{u_i}}{m_W \sin\beta} (O_N)_{A4} U^u_{X,i+3} \},
\nonumber \\
  N^{L(u)}_{iAX}&=& -\frac{g_2}{\sqrt{2}}
        \{ \frac{m_{u_i}}{m_W\sin\beta} (O_N)_{A4} U^u_{X,i}
         -\frac{4}{3} \tan\theta_W (O_N)_{A1} U^u_{X,i+3}  \}.
\end{eqnarray}

\section{Anomalous magnetic dipole-moment of the muon}
The magnetic dipole-moment interaction of muon is written as the following
form;
\begin{eqnarray}
\frac{ie}{2 m_\mu} F(q^2)\bar{u}(p_f) \sigma_{\mu \nu} q^{\mu} \epsilon^\nu
u(p_i),
\end{eqnarray}
where $q=p_f-p_i$ and $\epsilon$ the polarization vector of external
photon.  Then, the anomalous magnetic dipole-moment of muon is
\begin{eqnarray}
{(g-2)_\mu}\equiv 2 F(q^2=0).
\end{eqnarray}
We can write SUSY contributions
as $(g-2)_{\mu}^{SUSY}=(g^{(C)}+g^{(N)})_{\mu}$.
The first term $g^{(C)}_\mu$ represents the chargino-loop contribution as
\begin{eqnarray}
\nonumber
g^{(C)}_{\mu}&=&\frac{1}{48 \pi^2} \frac{m_{\mu}^2}{m_{\tilde{\nu}_X}^2}
|C_{2AX}^{L(l)}|^2
\frac{2+3 x_{AX}-6 x^2_{AX}+x^3_{AX} + 6x_{AX} \ln{x_{AX}}}
{(1-x_{AX})^4}
\\ \nonumber
&&+\frac{1}{16 \pi^2} \frac{m_{\mu}M_{\tilde{\chi}_A^-}}
{m_{\tilde{\nu}_X}^2}
C_{2AX}^{L(l)} C_{2AX}^{R(l)*}
\frac{-3 +4x_{AX}- x^2_{AX}-2 \ln{x_{AX}}}{(1-x_{AX})^3}
\\ &&+( L \leftrightarrow R)
\end{eqnarray}
where $x_{AX}=M_{\tilde{\chi}_A^-}^2/m_{\tilde{\nu}_X}^2$.

The neutralino-loop contribution $g^{(N)}_{\mu}$ is
\begin{eqnarray}
\nonumber
g^{(N)}_{\mu}&=&-\frac{1}{48 \pi^2} \frac{m_{\mu}^2}
{m_{\tilde{l}_X}^2}|N_{2AX}^{L(l)}|^2
\frac{1-6 x_{AX}+3 x^2_{AX}+2x^3_{AX} - 6x_{AX}^2 \ln{x_{AX}}}
{(1-x_{AX})^4}
\\ \nonumber
&&-\frac{1}{16 \pi^2} \frac{m_{\mu}M_{\tilde{\chi}_A^0}}
{m_{\tilde{l}_X}^2}
N_{2AX}^{L(l)} N_{2AX}^{R(l)*}
\frac{1-x_{AX}^2+2 x_{AX}\ln{x_{AX}}}{(1-x_{AX})^3}
\\ &&+(L \leftrightarrow R),
\end{eqnarray}
where $x_{AX}=M_{\tilde{\chi}_A^0}^2/m_{\tilde{l}_X}^2$.

\newpage
%
%
\newcommand{\Journal}[4]{{\sl #1} {\bf #2} {(#3)} {#4}}
\newcommand{\APJ}{Ap. J.}
\newcommand{\CJP}{Can. J. Phys.}
\newcommand{\NC}{Nuovo Cimento}
\newcommand{\NP}{Nucl. Phys.}
\newcommand{\PL}{Phys. Lett.}
\newcommand{\PR}{Phys. Rev.}
\newcommand{\PRep}{Phys. Rep.}
\newcommand{\PRL}{Phys. Rev. Lett.}
\newcommand{\PTP}{Prog. Theor. Phys.}
\newcommand{\SJNP}{Sov. J. Nucl. Phys.}
\newcommand{\ZP}{Z. Phys.}

\newpage

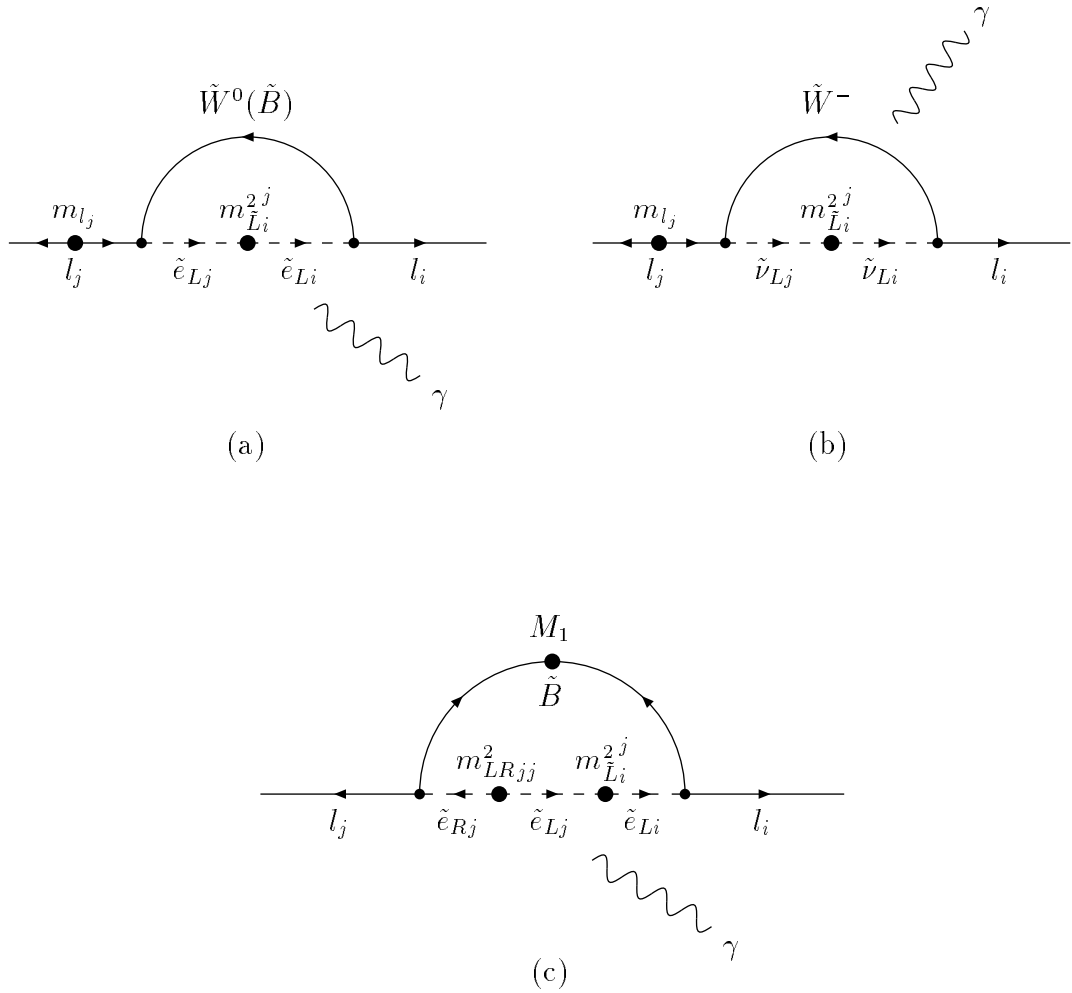
\begin{figure}
\epsfxsize=15cm
\caption
{Feynman diagrams which give rise to $l_j\rightarrow l_i\gamma$.  The
symbols $\tilde{e}_{Li}$, $\tilde {\nu}_{Li}$, $\tilde{B}$,
$\tilde{W}^{0}$, and $\tilde{W}^{-}$ represent left-handed charged
sleptons, left-handed sneutrinos, bino, neutral wino, and charged
wino, respectively.  In (a) and (b), the blob in the the
slepton/sneutrino line indicates the flavor-violating mass insertion
of the left-handed slepton and another blob in the external line the
chirality flip of the external lepton $l_j$. In (c), the blobs in
the slepton line indicate the insertions of the flavor-violating mass
(${m_{\tilde L}^2}^j_i$) and the left-right mixing mass
($m_{LRjj}^2$), and another blob in the bino line the chirality flip
of the bino $\tilde B$.}
\end{figure}
%
%
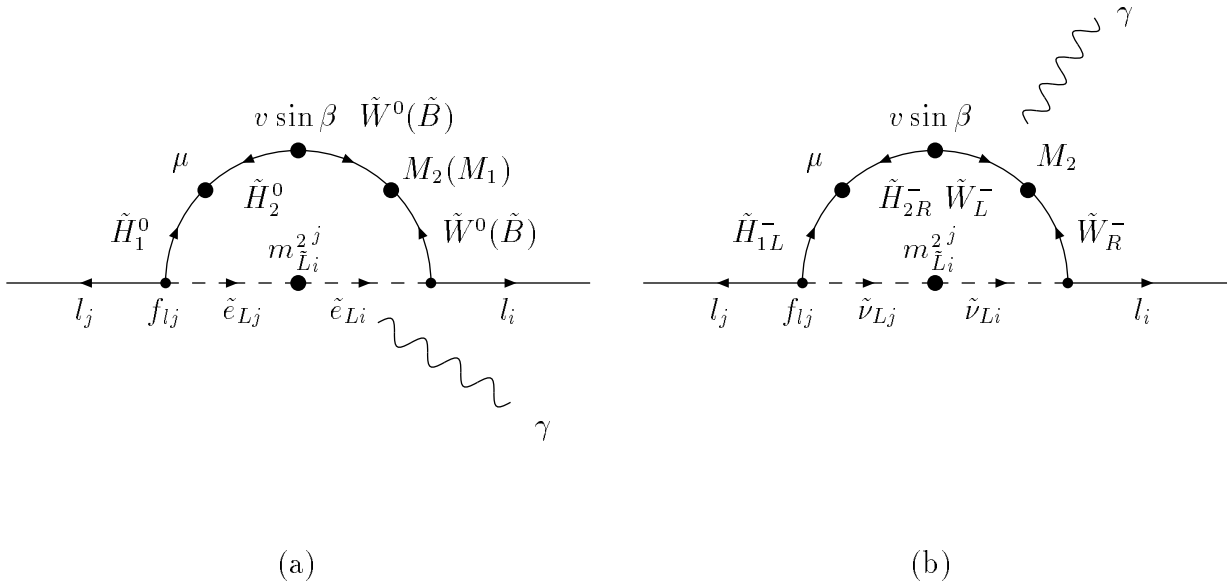
\begin{figure}[p]
\epsfxsize=15cm
\caption
{Feynman diagrams which give rise to the large $\tan \beta$
enhancement due to the gaugino-higgsino mixing in the process of
$l_j\rightarrow l_i\gamma$.  The symbols $\tilde{e}_{Li}$, $\tilde
{\nu}_{Li}$, $\tilde{B}$, $\tilde{W}^{0}$, $\tilde{W}^{-}$,
$\tilde{H}^{0}$, and $\tilde{H}^{-}$ represent left-handed charged
sleptons, left-handed sneutrinos, bino, neutral wino, charged wino,
neutral higgsino, and charged higgsino, respectively.  The blob in the
slepton/sneutrino line indicates the insertion of the flavor-violating
mass (${m_{\tilde L}^2}_i^j$). The blobs in the gaugino-higgsino line
indicate the mass insertions for gaugino-higgsino mixing, that is,
$\mu$ denotes higgsino ($\tilde{H}_1$-$\tilde{H}_2$) mass mixing,
$v \sin \beta$ the gaugino-higgsino ($\tilde{H}_2$-$\tilde{W}$)
mass mixing, and $M_2$ the wino mass. The value of $\tan \beta$ comes
from Yukawa coupling constant $f_{l_j} \sim 1/\cos \beta$ and v.e.v.
of $h_2$, $v \sin \beta$.}
\end{figure}
%
%
%
%
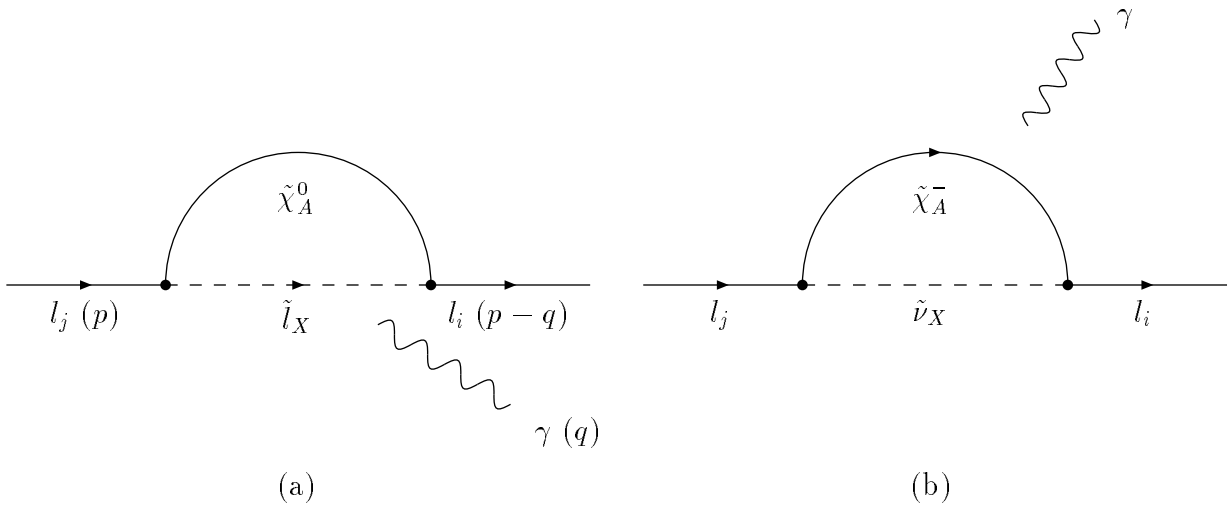
\begin{figure}[p]
\epsfxsize=15cm
\caption
{Feynman diagrams for the process $l_j \rightarrow l_i\gamma$.
(a) represents the contributions from neutralino
$\tilde{\chi}_A^0$ and slepton $\tilde{l}_X$ loop, and (b) the
contributions from chargino $\tilde{\chi}_A^-$ and sneutrino
$\tilde{\nu}_X$ loop. }
\end{figure}
%
%
%
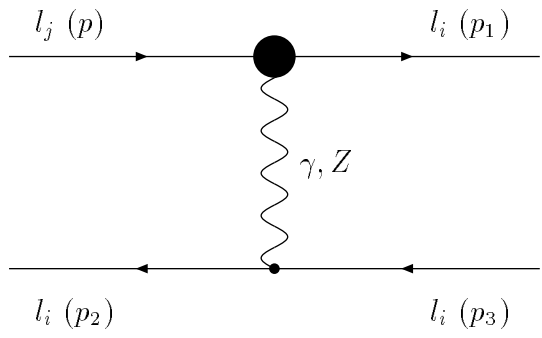
\begin{figure}[p]
\epsfxsize=15cm
\caption{Penguin-type diagrams for the process
$l_j^- \rightarrow l_i^- l_i^- l_i^+$ in which photon $\gamma$ and
$Z$-boson are exchanged.  The blob indicates $l_j$-$l_i$-$\gamma$
vertex such as Fig.3 or $l_j$-$l_i$-$Z$ vertex where $Z$-boson is
external.}
\end{figure}

%
%
%
%
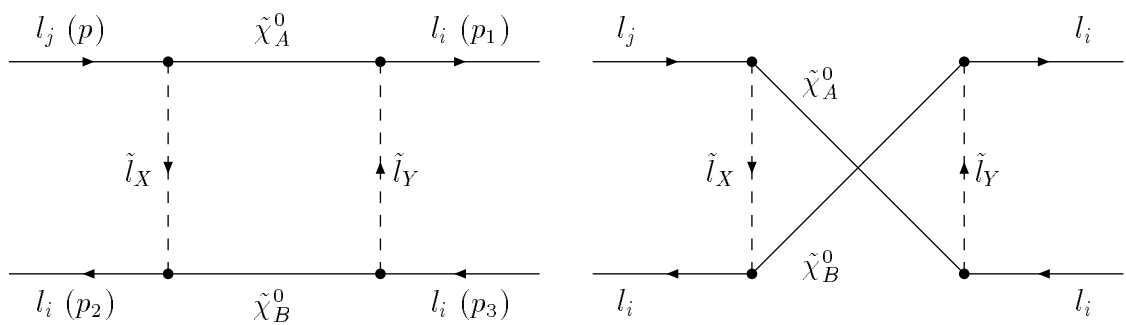
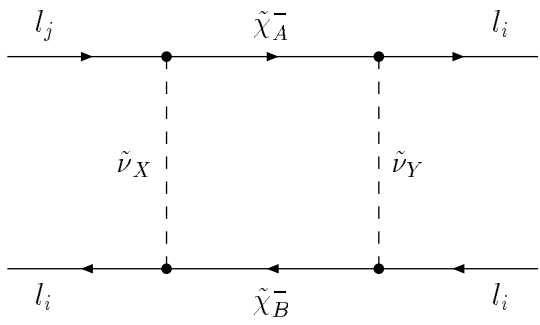
\begin{figure}[p]
\epsfxsize=15cm
\caption
{Box-type diagrams for the process $l_j^- \rightarrow l_i^- l_i^-
l_i^+$. Here, (a) represents the contributions from neutralino
$\tilde{\chi}_A^0$ and slepton $\tilde{l}_X$ loop, while (b) the
contributions from chargino $\tilde{\chi}_A^-$ and sneutrino
$\tilde{\nu}_X$ loop.}
\end{figure}
%
%
%
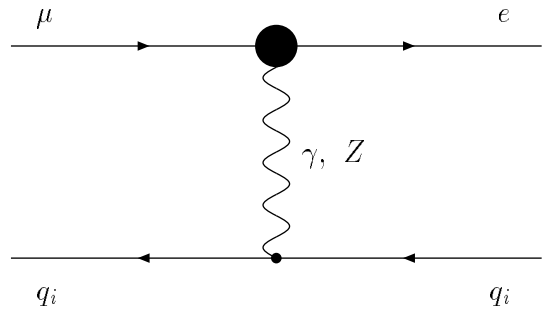
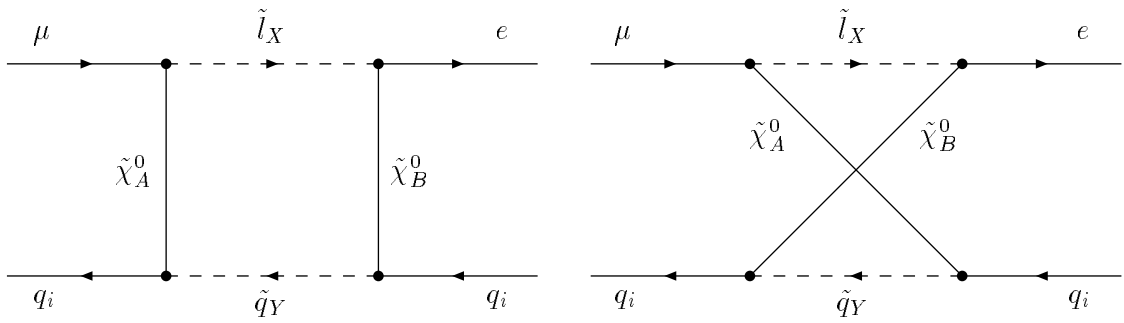
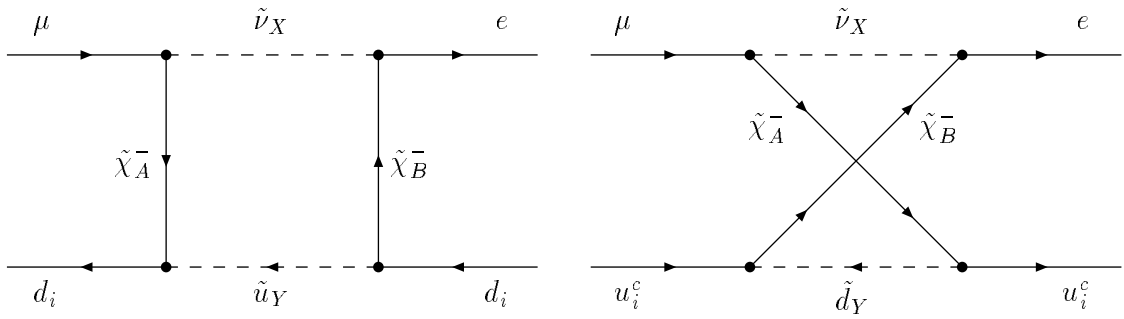
\begin{figure}[p]
\epsfxsize=15cm
\caption
{Feynman diagrams for the process $\mu$-$e$ conversion at the quark
level.  In (a), the Penguin-type diagram is depicted. The blob
indicates $l_j$-$l_i$-$\gamma$ vertex such as Fig.~3 or
$l_j$-$l_i$-$Z$ vertex such as Fig.~4. In (b) and (c), the box-type
diagrams are depicted; {\it i.e.}, (b) represents the contributions
from neutralino $\tilde{\chi}_A^0$, slepton $\tilde{l}_X$ and squark
$\tilde{q}_X~~(q=u,d)$ loop, and (c) the contributions from chargino
$\tilde{\chi}_A^-$, sneutrino $\tilde{\nu}_X$ and squark
$\tilde{q}_X~~(q=u,d)$ loop.}
\end{figure}
%
%
%
%
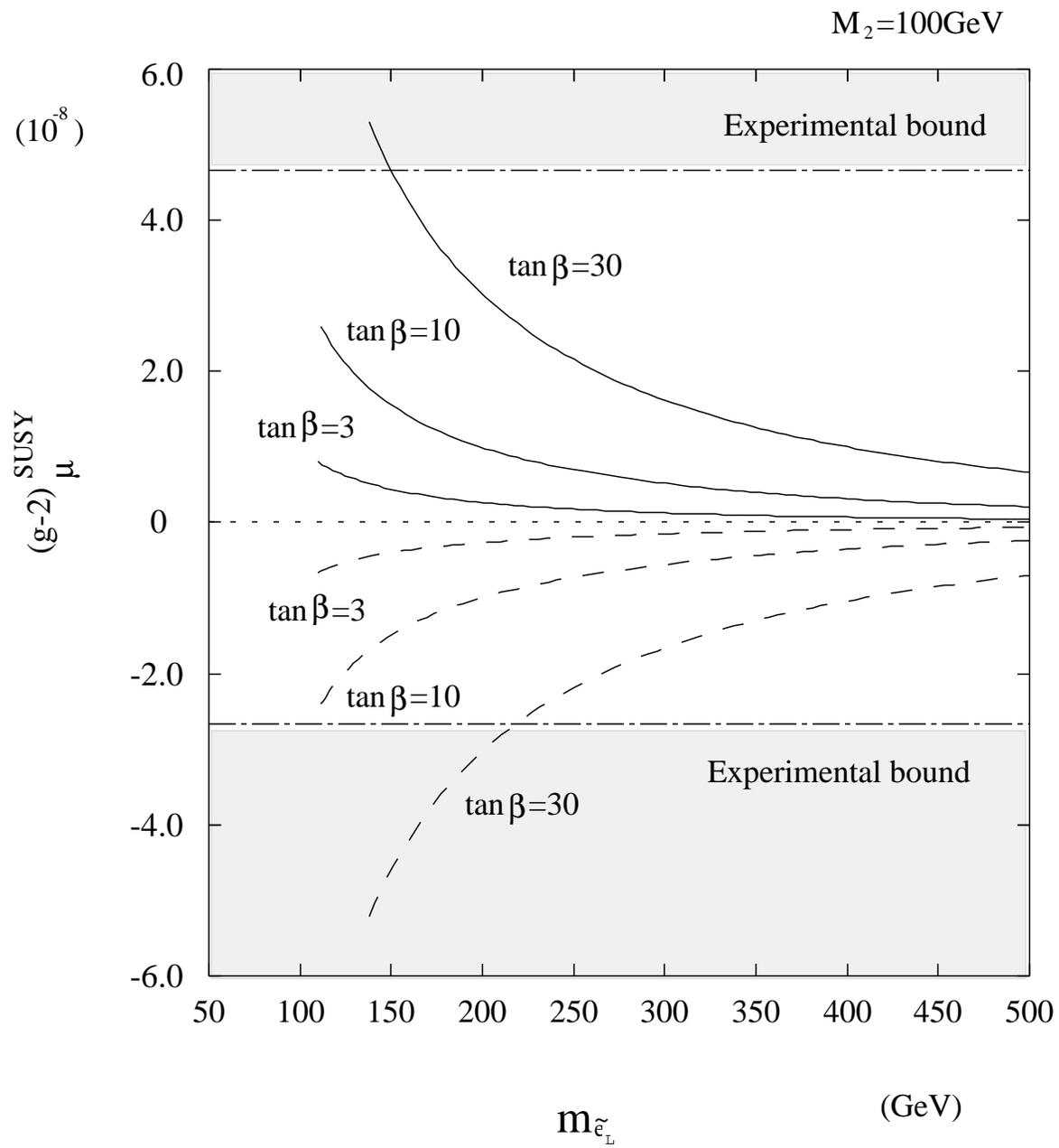
\begin{figure}[p]
\epsfxsize=15cm
\caption{The values of the SUSY contribution to the anomalous magnetic
dipole-moment of muon $(g-2)_\mu^{SUSY}$ as a function of the
left-handed selectron mass with the D-term contribution, which we
denote by $m_{{\tilde e}_L}$. Here we assume $a=0$ at the
gravitational scale. Real lines correspond to the case for $\mu>0$,
while dashed lines for $\mu<0$.  Here we have taken $M_2=100$ GeV and
$\tan\beta=3,~10,~30$. The shaded regions are excluded by the present
experiments.}
\end{figure}
%
%
%
%
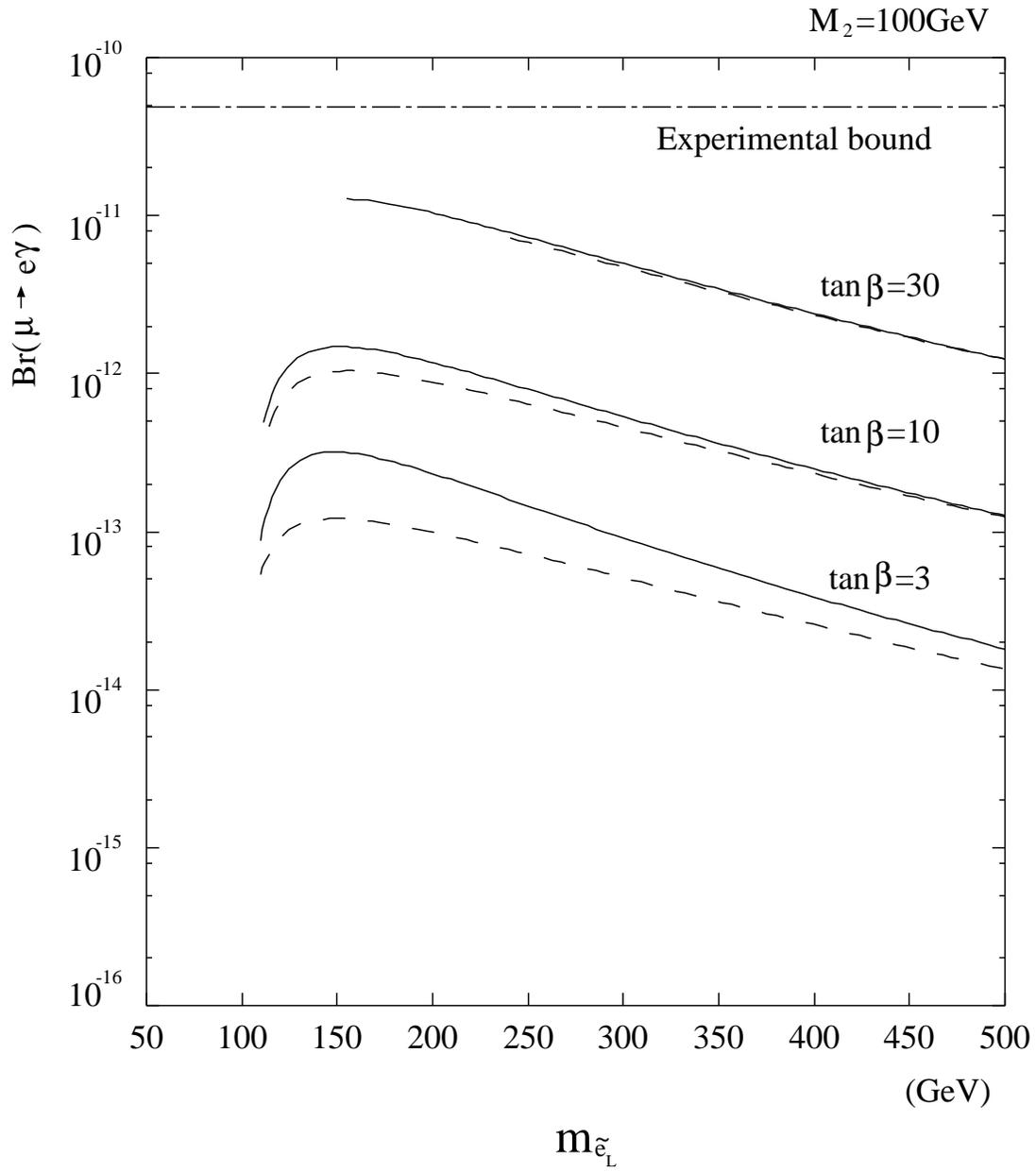
\begin{figure}[p]
\epsfxsize=15cm
\caption
{Branching ratios for the process $\mu \rightarrow e \gamma$ in the
{\em Case 1\/}) $V= V_{KM}$ as a function of the left-handed selectron
mass with the D-term contribution, $m_{{\tilde e}_L}$. Real lines
correspond to the case for $\mu>0$, while dashed lines for $\mu<0$.
Here we have taken $M_2=100$ GeV and $\tan \beta=3,~10,~30$. We also
show the present experimental upper bound for this process by the
dash-dotted line.}
\end{figure}
%
%
%
%
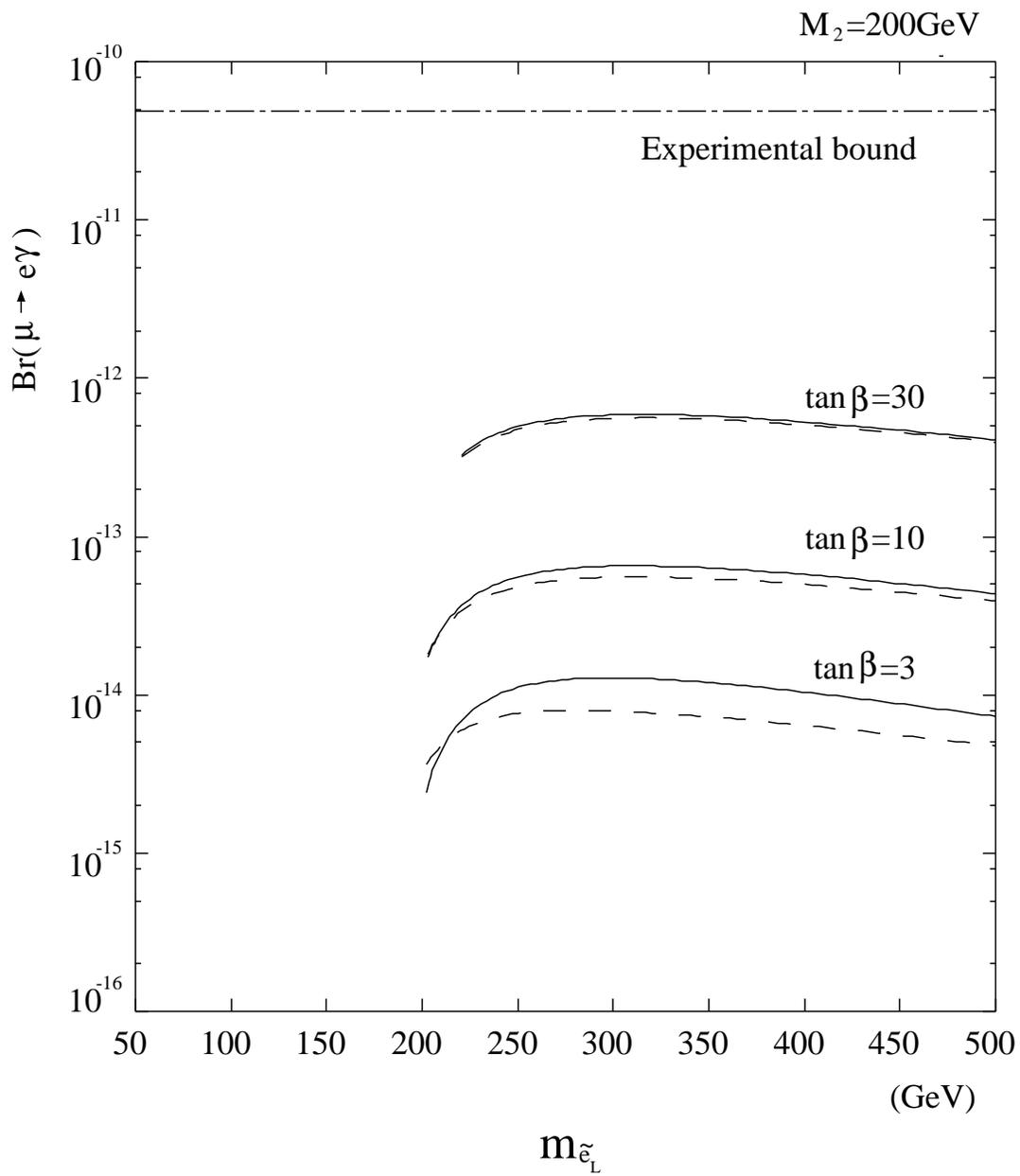
\begin{figure}[p]
\epsfxsize=15cm
\caption{Same as Fig.~8 except for $M_2=200$ GeV.}
\end{figure}
%
%
%
%
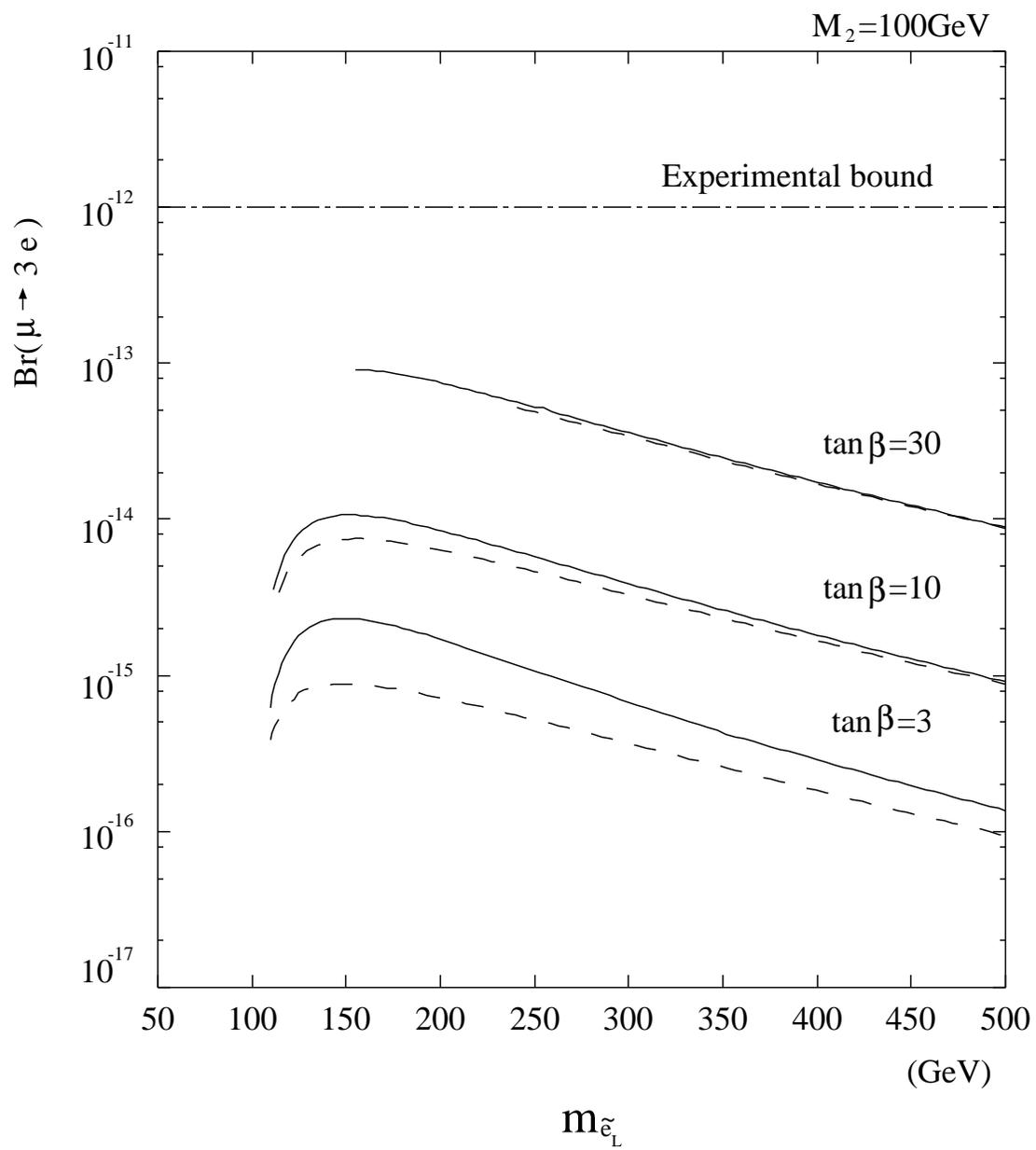
\begin{figure}[p]
\epsfxsize=15cm
\caption
{Branching ratios for the process $\mu^- \rightarrow e^- e^- e^+$ in
the {\em Case 1\/}) $V= V_{KM}$ as a function of the left-handed
selectron mass with the D-term contribution, $m_{{\tilde e}_L}$. Real
lines correspond to the case for $\mu>0$, while dashed lines for
$\mu<0$. Here we have taken $M_2=100$ GeV and $\tan \beta=3,~10,~30$.
We also show the present experimental upper bound for this process by
the dash-dotted line.}
\end{figure}
%
%
%
%
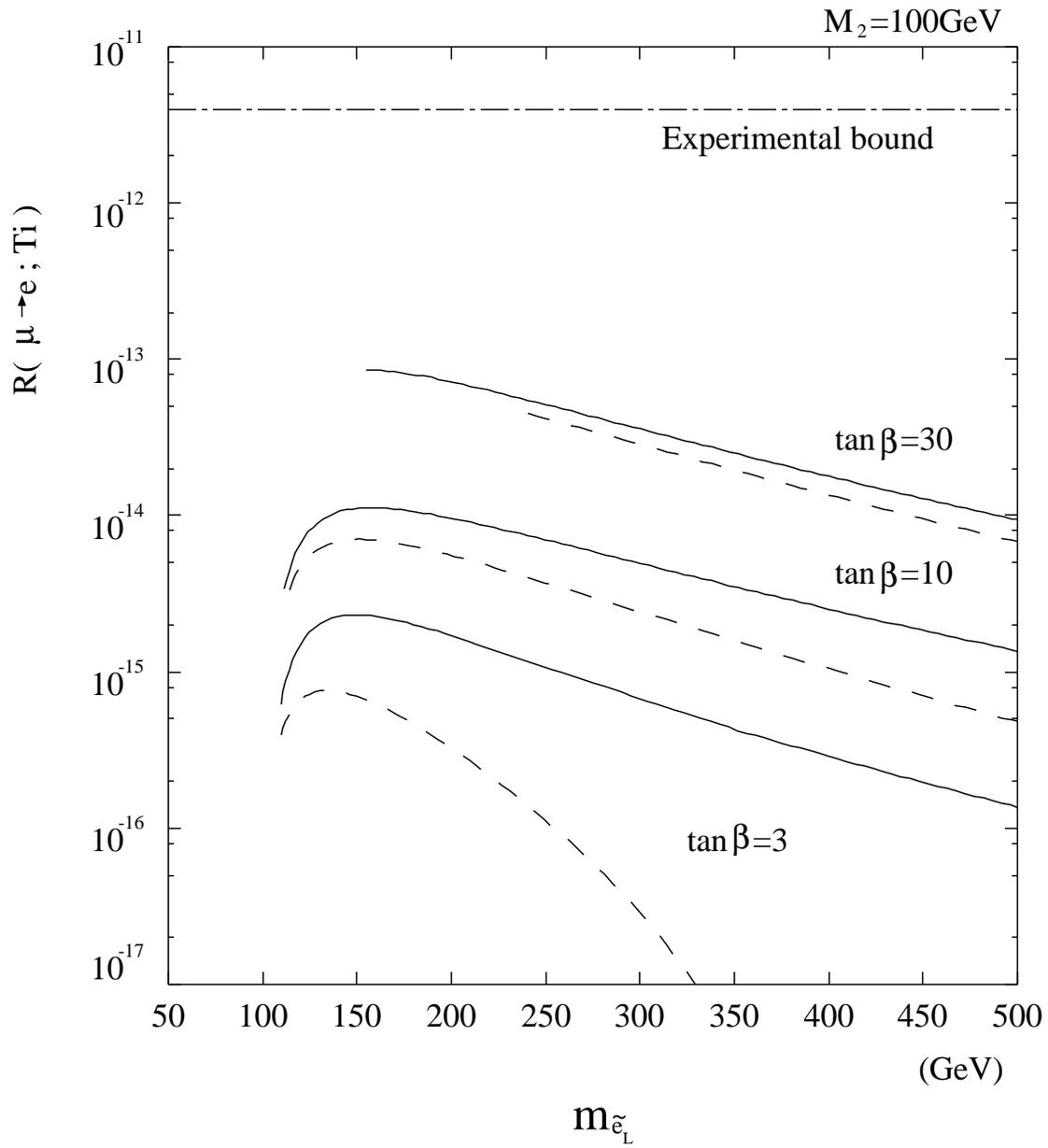
\begin{figure}[p]
\epsfxsize=15cm
\caption
{The $\mu$-$e$ conversion rates in nuclei $^{48}_{22}{\rm Ti}$ in the
{\em Case 1\/}) $V= V_{KM}$ as a function of the left-handed selectron
mass with the D-term contribution, $m_{{\tilde e}_L}$. Real lines
correspond to the case for $\mu>0$, while dashed lines for $\mu<0$.
Here we have taken $M_2=100$ GeV and $\tan \beta=3,~10,~30$. We also
show the present experimental upper bound for this process by the
dash-dotted line.}
\end{figure}
%
%
%
%
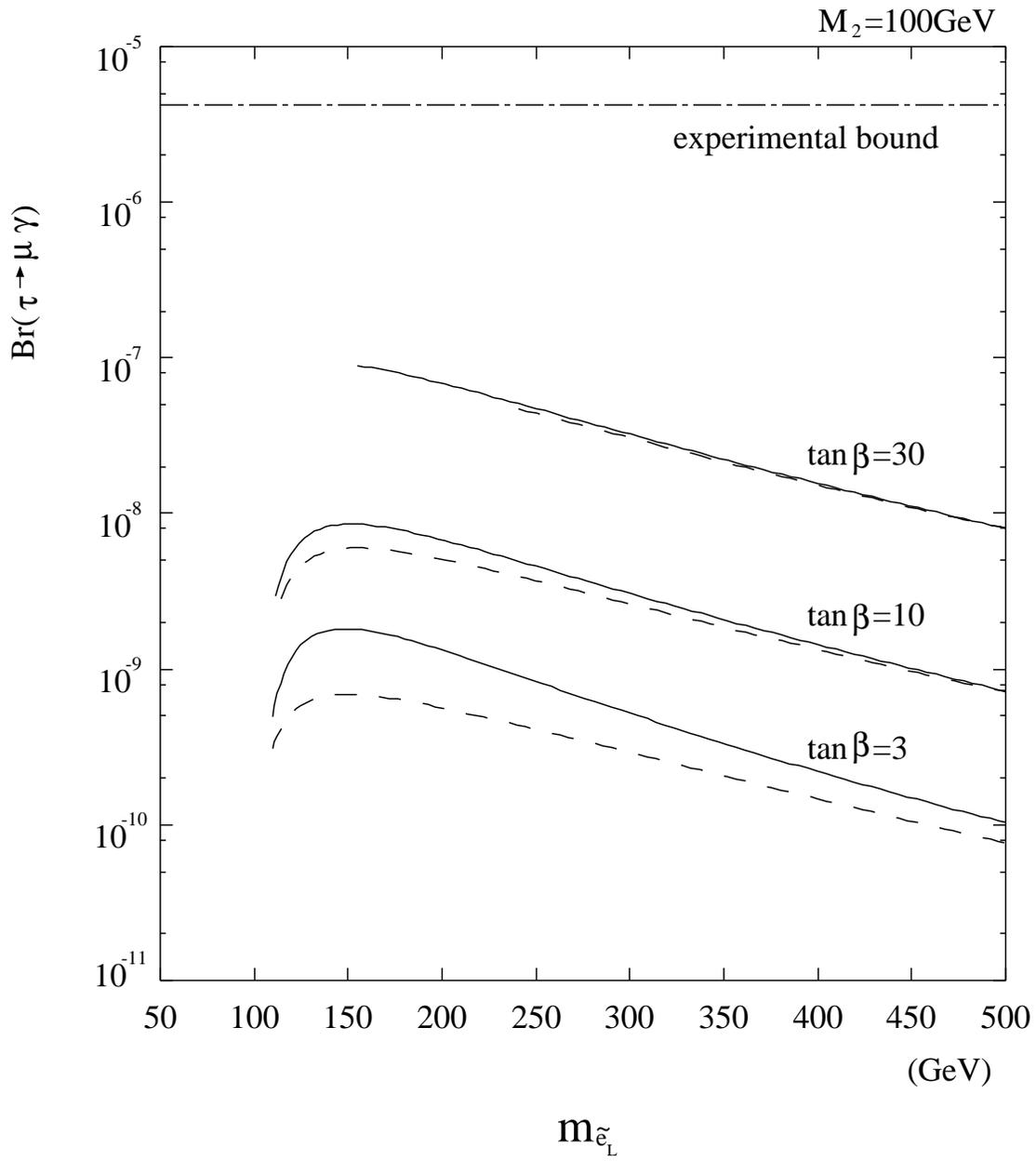
\begin{figure}[p]
\epsfxsize=15cm
\caption
{Branching ratios for the process $\tau \rightarrow \mu \gamma$ in the
{\em Case 1\/}) $V= V_{KM}$ as a function of the left-handed selectron
mass with the D-term contribution, $m_{{\tilde e}_L}$. Real lines
correspond to the case for $\mu>0$, while dashed lines for $\mu<0$.
Here we have taken $M_2=100$ GeV and $\tan \beta=3,~10,~30$. We also
show the present experimental upper bound for this process by the
dash-dotted line.}
\end{figure}
%
%
%
%
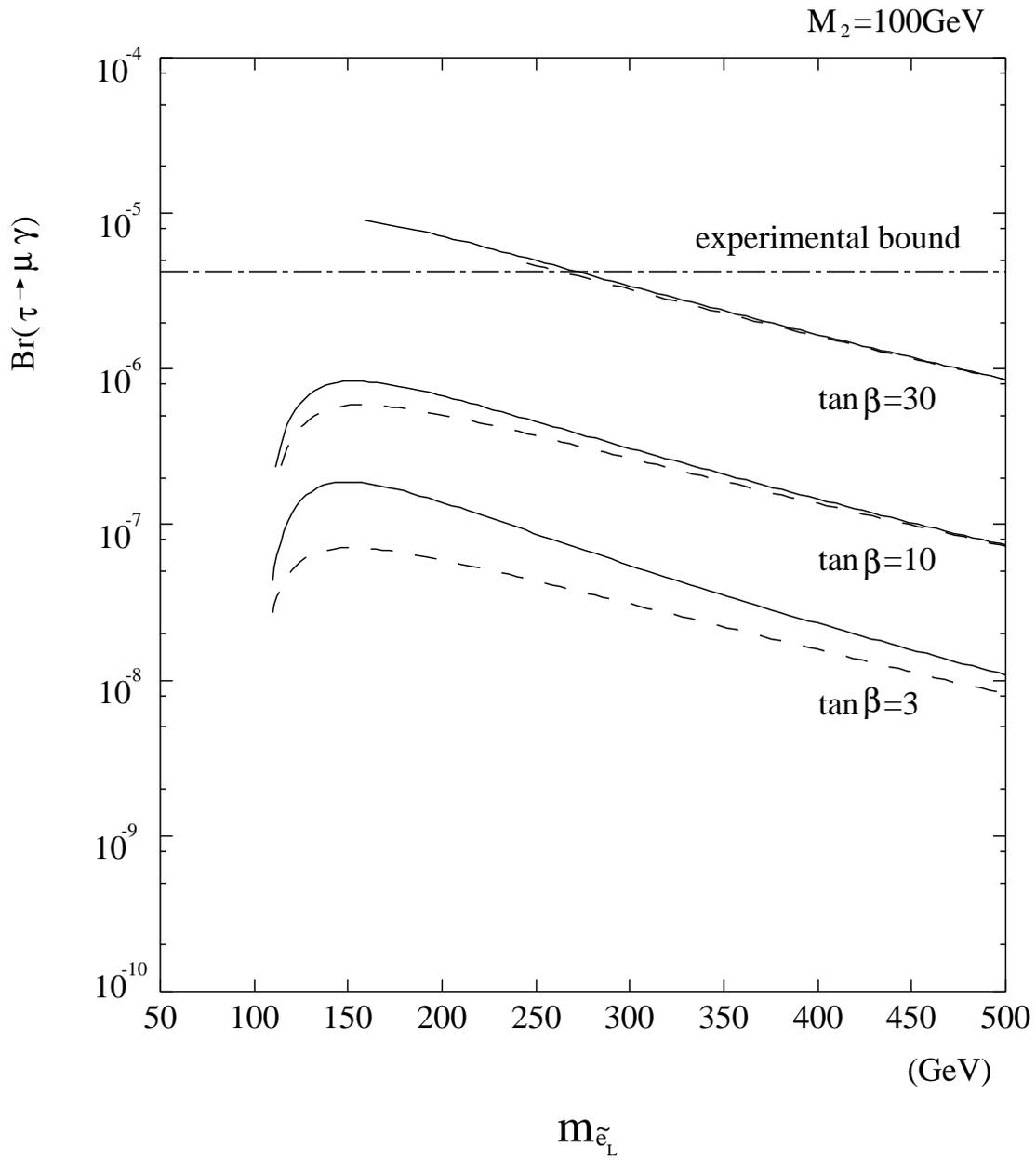
\begin{figure}[p]
\epsfxsize=15cm
\caption
{Branching ratios for the process $\tau \rightarrow \mu \gamma$ in the
{\em Case 2\/}) neutrino mixing implied by atmospheric neutrino
deficit, as a function of the left-handed selectron mass with the
D-term contribution, $m_{{\tilde e}_L}$. Real lines correspond to the
case for $\mu>0$, while dashed lines for $\mu<0$. Here we have taken
$M_2=100$ GeV and $\tan\beta=3,~10,~30$. We also show the present
experimental upper bound for this process by the dash-dotted line.}
\end{figure}
%
%
%
%
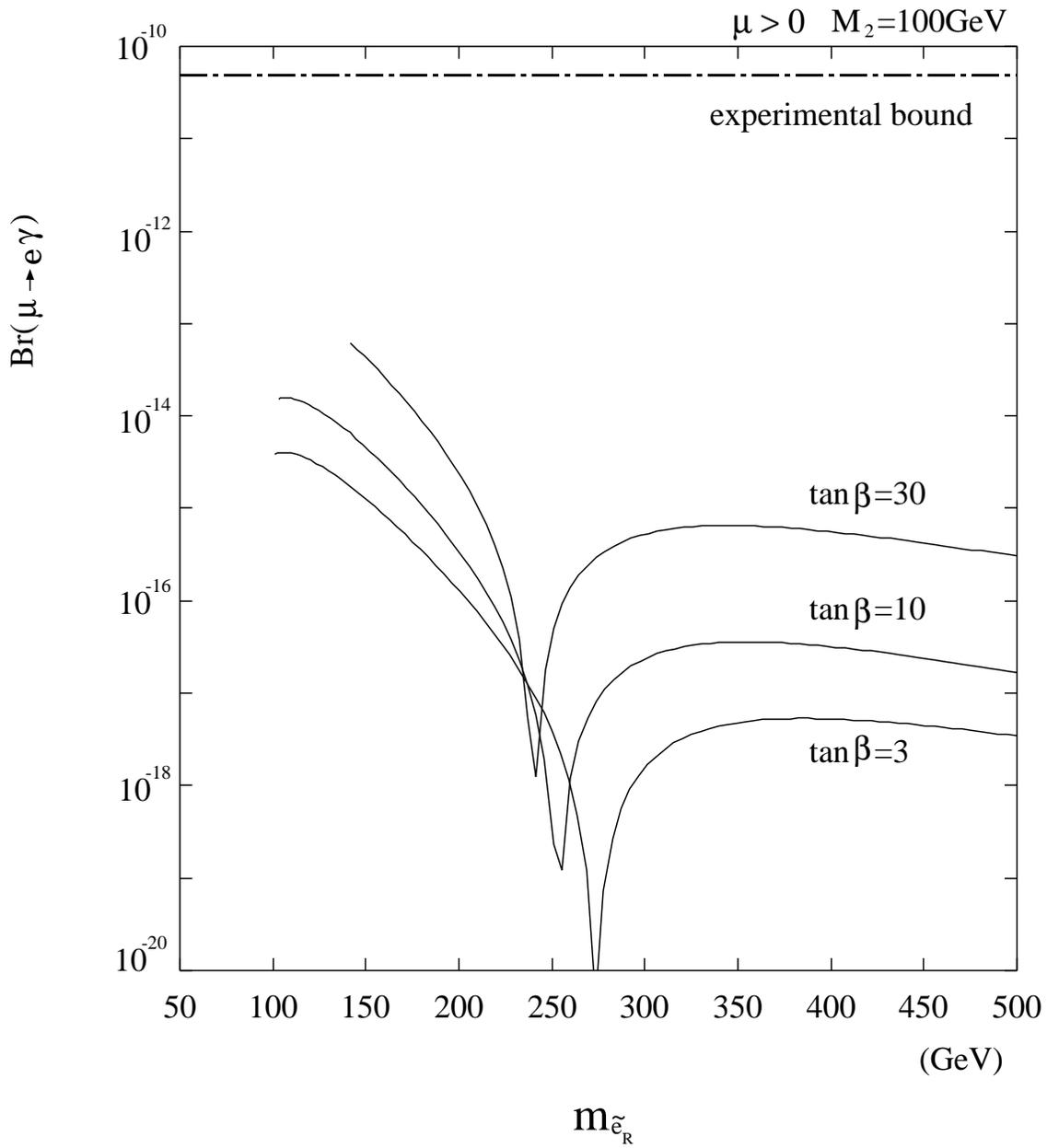
\begin{figure}[p]
\epsfxsize=15cm
\caption
{Branching ratios for the process $\mu \rightarrow e \gamma$ in the
case for the minimal $SU(5)$ grand unified theory, as a function of
the right-handed selectron mass with the D-term contribution,
$m_{{\tilde e}_R}$.  Here we have taken $\mu>0$, $M_2=100$ GeV, and
$\tan\beta=3,~10,~30$. We also show the present experimental upper
bound for this process by the dash-dotted line.}
\end{figure}

\end{document}